\documentclass[english,aps,twocolumn,groupedaddress,pra,superscriptaddress,amsmath,amssymb,noeprint,nolongbibliography]{revtex4-2}
\usepackage[T1]{fontenc}
\usepackage{babel}
\usepackage{float}
\usepackage{mathrsfs}
\usepackage{amsmath}
\usepackage{amssymb}
\usepackage{graphicx}
\usepackage{esint}
\usepackage[dvipsnames]{xcolor}
\usepackage{soul}
\usepackage{ulem,xpatch}
\usepackage{comment}
\usepackage[braket, qm]{qcircuit}
\usepackage{mleftright} % For adequate spacing in formulas
\usepackage[unicode=true,
 bookmarks=false,
 breaklinks=false,pdfborder={0 0 1},backref=section,colorlinks=false]
 {hyperref}



\begin{document}
\title{Optimized continuous dynamical decoupling via differential geometry and machine learning}

\author{N\'icolas Andr\'e da Costa Morazotti}
\affiliation{S\~ao Carlos Institute of Physics, University of S\~ao Paulo, PO Box 369,
13560-970, S\~ao Carlos, SP, Brazil}
\author{Adonai Hil\'ario da Silva}
\affiliation{S\~ao Carlos Institute of Physics, University of S\~ao Paulo, PO Box 369,
13560-970, S\~ao Carlos, SP, Brazil}
\author{Gabriel Audi}
\affiliation{S\~ao Carlos Institute of Physics, University of S\~ao Paulo, PO Box 369,
13560-970, S\~ao Carlos, SP, Brazil}
\author{Felipe Fernandes Fanchini}
\affiliation{São Paulo State University (UNESP), School of
Sciences, 17033-360, Bauru, SP, Brazil}
\affiliation{QuaTI - Quantum Technology \& Information, 13560-161, São Carlos, SP, Brazil}
\author{Reginaldo de Jesus Napolitano}
\email{reginald@ifsc.usp.br}
\affiliation{S\~ao Carlos Institute of Physics, University of S\~ao Paulo, PO Box 369,
13560-970, S\~ao Carlos, SP, Brazil}

\begin{abstract}
We introduce a strategy to develop optimally designed fields for continuous dynamical decoupling (CDD). Our methodology obtains the optimal continuous field configuration to maximize the fidelity of a general one-qubit quantum gate. To achieve this, considering dephasing-noise perturbations, we employ an auxiliary qubit instead of the boson bath to implement a purification scheme, which results in unitary dynamics. Employing the sub-Riemannian geometry framework for the two-qubit unitary group, we derive and numerically solve the geodesic equations, obtaining the optimal time-dependent control Hamiltonian. Also, due to the extended time required to find solutions to the geodesic equations, we train a neural network on a subset of geodesic solutions, enabling us to promptly generate the time-dependent control Hamiltonian for any desired gate, which is crucial in circuit optimization.
\end{abstract}

\maketitle

\section{Introduction}

The burgeoning investment in quantum computing devices will likely
allow the first steps to transcending the current noisy
intermediate-scale quantum (NISQ) paradigm~\cite{RevModPhys.94.015004}.
Indeed even during the NISQ era~\cite{Preskill2018quantumcomputingin},
we already see signs of a revival of interest in practical quantum
error correction~\cite{Acharya2023}. While several error-correction schemes are rapidly developing, we also highlight the relevance of error-suppression methodologies. Among these approaches, dynamical decoupling is noteworthy for a significant property: it does not require auxiliary qubits for protection, making it a promising approach to this fast-evolving landscape. Consequently, the relevance of quantum gate control at the pulse level~\cite{9835639}, within the dynamical decoupling time regime, becomes apparent.  The study of continuous dynamical decoupling (CDD) has seen significant advancements in recent years. Notably, other researchers in the field have explored robust dynamical decoupling with concatenated continuous driving, demonstrating the potential for protecting quantum coherence in various systems~\cite{Cai_2012}. Additionally, applying these techniques to protect quantum spin coherence in nanodiamonds within living cells further highlights the practical implications of this approach~\cite{PhysRevApplied.13.024021}.

It is our view that the adoption of the current NISQ concept, which assumes a fixed set of universal quantum gates, is not as flexible as the ability to generate local gates that, optimally, satisfy the intended algorithm at each particular circuit location. According to an optimal schedule, taking into account the perturbations known to be present at the circuit site, the local gates must be optimally controlled so that their output is delivered at the right time with the required fidelity. Hence, given a desired algorithm and the characteristics of the hardware noise, previously characterized, each gate in the circuit, as well as the complete circuit, is to be optimized. We start investigating such a new quantum computation paradigm by optimizing only the control of single-qubit gates. These gates are required on demand to deliver output at a specific time, despite the presence of perturbations, scheduled by the algorithm. In the service of the algorithm design, depending on the circuit site and specifics of the intended computation, the optimized single-qubit gates must be rendered efficiently on demand. Our solution is to have a trained neural network to expedite the estimated control field for any required single-qubit gate such a design demands, as we show in the following.

A quantum circuit consists of a set of
universal quantum gates~\cite{Nielsen2010-sk}. A simple
universal set of quantum gates requires a two-qubit entangling interaction
and arbitrary single-qubit operations. In the absence of noise, even
if the two-qubit interaction can not be externally controlled, by
controlling the single-qubit gates instead, with such a set it is
possible to effectively produce any two-qubit gate on demand, following
a precise schedule. In the presence of noise, the ability to externally
operate on single qubits must implement the intended target operation
despite the noisy perturbations. Single-qubit control is, therefore,
crucial to building high-fidelity elementary one- and two-qubit modules~\cite{PhysRevA.89.022317,Kannan2023}
to operate as logical gates in a quantum circuit.

A quantum circuit might be subject to several kinds
of quantum perturbations, but the dephasing of memory qubits can be
solved analytically and is one of the most critical errors because
it destroys quantum superposition, ruining such an important resource
essential to quantum computing~\cite{Nielsen2010-sk,PhysRevA.51.992}.
During a single-qubit operation in the presence of dephasing noise,
however, dissipation might also occur besides decoherence, depending
on the operation being performed. Hence, we will consider dephasing noise and amplitude damping noise, since, in general, all other perturbations will, effectively, result in a combination of these two basic disturbances. Our purpose is thus to calculate
the optimal time-dependent control Hamiltonian resulting in a desired high-fidelity gate operation. Moreover, the single-qubit control must be optimized to minimize a cost function (energy consumption) during a time interval, despite the perturbations. As we will show, the control Hamiltonian 
satisfying these conditions is a geodesic in a geometry whose metric is introduced by the cost function.

Despite the promising framework, obtaining a target geodesic is computationally time-consuming~\cite{PhysRevLett.114.170501}. To solve this problem, after calculating some target geodesics, we iteratively use a type of physics-informed neural network to predict the time-dependent Hamiltonian associated with the desired gate operation. The neural network guesses are then used as the new seeds in the minimization protocol, decreasing the optimization time. As more data are generated, the neural network returns increasingly better guesses, since the process can be repeated until a quality threshold is achieved. Our method can produce the desired quantum gate required by the global circuit. It is important to emphasize that our approach addresses dephasing noise during the operation of any single-qubit gate, which includes dissipation depending on the gate required. Still, a similar scheme can be used to curb other kinds of noise, as we illustrate by showing how to adapt our procedure to the case of amplitude damping. For other kinds of perturbations, an alternative approach, resembling ours, has been presented recently~\cite{aroch_mitigating_2023}.

This paper is organized as follows. Section II presents the theory adopted and elucidates the process for calculating the optimal control field for a general quantum gate. Section III, to ensure efficient computation of the control field, presents the use of a neural network to extend the numerical simulation for real-time applications. Section IV presents the protection for Hadamard, $X,$ and $T$ gates, as well as the `Identity' gate, which can be thought as the process of keeping the state memory for the duration a quantum gate application, and Sec. V concludes with a summary of our results and an outline of our future investigations.

\section{Quantum Optimal Control}\label{methods}

An efficient method to protect a gate from environmental noise is the well-known continuous dynamical decoupling (CDD)~\cite{FONSECAROMERO2004307,PhysRevLett.95.140502,PhysRevA.73.022343,PhysRevA.75.022329, PhysRevD.107.014506, PhysRevA.106.042434, Napolitano2021protecting}. CDD can protect a general gate against uncorrelated errors without prior noise characterization. Although infinite external field configurations can decouple a gate from an arbitrary reservoir, some setups are distinctly more efficient, requiring less energy while still protecting the quantum system with high fidelity. The strategy we present in this manuscript is precisely oriented toward this end: identifying the optimal configuration that minimizes energy consumption during logical gate protection. For this purpose, when the noise we intend to mitigate has been characterized by some means, as described, for instance, in Refs.~\cite{Youssry2020,Szankowski2017, PhysRevLett.107.230501, PhysRevApplied.15.014033, PhysRevLett.107.170504, Almog_2011, Burgardt_2023, PhysRevLett.126.250507, PhysRevLett.111.093604}, we look for a protocol that penalizes using high-energy fields while optimizing the time dependence of the control fields. Naturally, implementing an optimal control strategy, considering open quantum systems requires a comprehensive understanding of the specific reservoir under consideration. Indeed, the field optimization depends on the nuances of the environment, such as spectral density, the reservoir's correlation time, and other factors.

To demonstrate our methodology, we organize the theory and strategy into five subsections. In subsection \ref{model} we describe our model based on the Caldeira-Leggett theory of quantum Brownian motion~\cite{CALDEIRA1983587}. In sequence,  in subsection \ref{qoc}, we detail the purification process for the system of interest, ensuring the global dynamics are unitary. In subsection \ref{form} we show how to describe our optimization problem using Lagrange multipliers, which are further calculated in subsection \ref{subsec:Procedure}. In subsection \ref{subsec:ampdamp} we show how this problem can be extended to amplitude-damping noise under specific conditions.

\subsection{Dephasing model}\label{model}

To illustrate our approach, we adopt a modified version of the spin-boson model of the dissipative
two-state system~\cite{RevModPhys.59.1}, as given
by the following total Hamiltonian:
\begin{align}
H_{\text{tot}}\mleft(t\mright) & = H_{c}\mleft(t\mright)+\sum_{\lambda}\hbar\omega_{\lambda}b_{\lambda}^{\dagger}b_{\lambda}\nonumber \\ & \phantom{{}={}} + \sigma_{z}\sum_{\lambda}\hbar\left(g_{\lambda}b_{\lambda}+g_{\lambda}^{\ast}b_{\lambda}^{\dagger}\right),\label{Htot(t)}
\end{align}
where
\begin{align}
H_{c}\mleft(t\mright) =  \hbar\omega_{x}\mleft(t\mright)\sigma_{x}+\hbar\omega_{y}\mleft(t\mright)\sigma_{y}+\hbar\omega_{z}\mleft(t\mright)\sigma_{z}\label{Hc(t)}
\end{align}
is the control Hamiltonian, $\sigma_{x},$ $\sigma_{y},$ and $\sigma_{z}$
are the Pauli matrices, $\omega_{x}\mleft(t\mright),$ $\omega_{y}\mleft(t\mright),$
and $\omega_{z}\mleft(t\mright)$ are time-dependent functions as a consequence of applied external control fields, as explained below,
$b_{\lambda}$ and $b_{\lambda}^{\dagger}$ are, respectively, the
annihilation and creation operators of the environmental boson in
mode $\lambda,$ and $g_{\lambda}$ is the coupling constant between
the qubit and the boson mode $\lambda.$ We regain the model Hamiltonian
of Ref.~\cite{RevModPhys.59.1} by assuming, for example, that when
the control fields are not present, the coefficients of the Pauli
matrices in Eq. (\ref{Hc(t)}) become $\omega_{x}\mleft(t\mright)=-\Delta/2,$
$\omega_{y}\mleft(t\mright)=0,$ and $\omega_{z}\mleft(t\mright)=\varepsilon/2,$
where $\Delta$ and $\varepsilon$ are constants.

It is convenient to use a unitary transformation given by $U_{B}\mleft(t\mright)=\exp\mleft(-it\sum_{\lambda}\omega_{\lambda}b_{\lambda}^{\dagger}b_{\lambda}\mright)$
and write the resulting total Hamiltonian in this new picture as
\begin{align}
H_{\text{new}}\mleft(t\mright) = H_{c}\mleft(t\mright)+\sigma_{z}B\mleft(t\mright), 
\label{Hnew(t)}
\end{align}
where we define the boson field emulating the environmental noise as
\begin{align}
B\mleft(t\mright) &\equiv \sum_{\lambda}\hbar\Bigl[g_{\lambda}b_{\lambda}\exp\mleft(-i\omega_{\lambda}t\mright)\nonumber \\
 & \phantom{{}={}} + g_{\lambda}^{\ast}b_{\lambda}^{\dagger}\exp\mleft(i\omega_{\lambda}t\mright)\Bigr].\label{B(t)}
\end{align}
The time-local, second-order master equation describing the evolution
of the reduced density matrix of the qubit, in the interaction picture,
is written as~\cite{Gordon_2007,PhysRevLett.93.130406,Chaturvedi1979,Shibata1977}
\begin{widetext}
\begin{align}
\frac{\mathrm{d}}{\mathrm{d}t}\rho_{IS}\mleft(t\mright) = -\frac{1}{\hbar^{2}}\int_{0}^{t}\mathrm{d}t^{\prime}\,\mathrm{Tr}_{B}\mleft\{ \left[H_{I}\mleft(t\mright),\left[H_{I}\mleft(t^{\prime}\mright),\rho_{B}\mleft(0\mright)\rho_{IS}\mleft(t\mright)\right]\right]\mright\} ,\label{master}
\end{align}\label{master_equation}
\end{widetext}
where $\rho_{IS}\mleft(t\mright)$ is the qubit reduced
density matrix, $\rho_{B}\mleft(0\mright)$ is the initial density matrix
of the environment represented by the boson bath, and the interaction-picture Hamiltonian is defined as
\begin{align}
H_{I}\mleft(t\mright) = S\mleft(t\mright)B\mleft(t\mright),\label{HI(t)}
\end{align}
with
\begin{align}
S\mleft(t\mright) = U_{S}^{\dagger}\mleft(t\mright)\sigma_{z}U_{S}\mleft(t\mright)\label{S(t)}
\end{align}
and $U_{S}\mleft(t\mright)$ is the unitary operator that satisfies
\begin{align}
i\hbar\frac{\mathrm{d}U_{S}\mleft(t\mright)}{\mathrm{d}t} = H_{c}\mleft(t\mright)U_{S}\mleft(t\mright)\label{dUS(t)dt}
\end{align}
and $U_{S}\mleft(0\mright) = I$.

When solving Eq. (\ref{master}), usually the initial state of the thermal bath is taken as the mixed state given by the canonical density-matrix
operator
\begin{align}
\rho_{B}\mleft(0\mright) = \frac{\exp\mleft(-\beta\hbar\sum_{s}\omega_{s}b_{s}^{\dagger}b_{s}\mright)}{Z},\label{thermal}
\end{align}
where $\beta=1/k_{B}T,$ $T$ is the boson-bath temperature, $k_{B}$
is Boltzmann constant, and $Z=\mathrm{Tr}_{B}\mleft[\exp\mleft(-\beta\hbar\sum_{s}\omega_{s}b_{s}^{\dagger}b_{s}\mright)\mright]$
is the partition function. In our treatment of pure dephasing noise
we simplify the structure of the noise by assuming an Ohmic spectral
density~\cite{gardiner_quantum_2004}, using
\begin{align}
J\mleft(\omega\mright) = \eta\omega\exp\mleft(-\omega/\omega_{c}\mright),\label{Ohmic}
\end{align}
where $\omega_{c}$ is a cutoff frequency and $\eta$ is a dimensionless
noise strength. In the absence of control and if in this case $H_{c}\mleft(t\mright)\propto\sigma_{z},$
Eq. (\ref{master}) has the analytical solution~\cite{RevModPhys.59.1,PhysRevA.65.032326,Breuer2007-ti}
\begin{align}
\rho_{IS}^{\text{no control}}\mleft(t\mright) = \left(\begin{array}{cc}
\left|c_{1}\right|^{2} & c_{1}c_{2}^{\ast}\mu\mleft(t\mright)\\
c_{1}^{\ast}c_{2}\mu\mleft(t\mright) & \left|c_{2}\right|^{2}
\end{array}\right),\label{exact}
\end{align}
where 
\begin{align}
\mu\mleft(t\mright) \equiv \left\{ \frac{\left|\left(\frac{k_{B}T}{\hbar\omega_{c}}+i\frac{k_{B}T}{\hbar}t\right)!\right|^{4}}{\left(1+\omega_{c}^{2}t^{2}\right)\left[\left(\frac{k_{B}T}{\hbar\omega_{c}}\right)!\right]^{4}}\right\} ^{2\eta},\label{decaying coherence}
\end{align}
and initial state described by $\rho_{IS}\mleft(0\mright)=\left|\psi\mleft(0\mright)\right\rangle \left\langle \psi\mleft(0\mright)\right|,$
with $\left|\psi\mleft(0\mright)\right\rangle =c_{1}\left|1\right\rangle +c_{2}\left|2\right\rangle $
and $\left|c_{1}\right|^{2}+\left|c_{2}\right|^{2}=1.$ Here we use
the factorial notation for the gamma function~\cite{georgearfken2011},
namely, $z!=\Gamma\mleft(z+1\mright)$.

\begin{figure}
\includegraphics[width=0.53\textwidth]{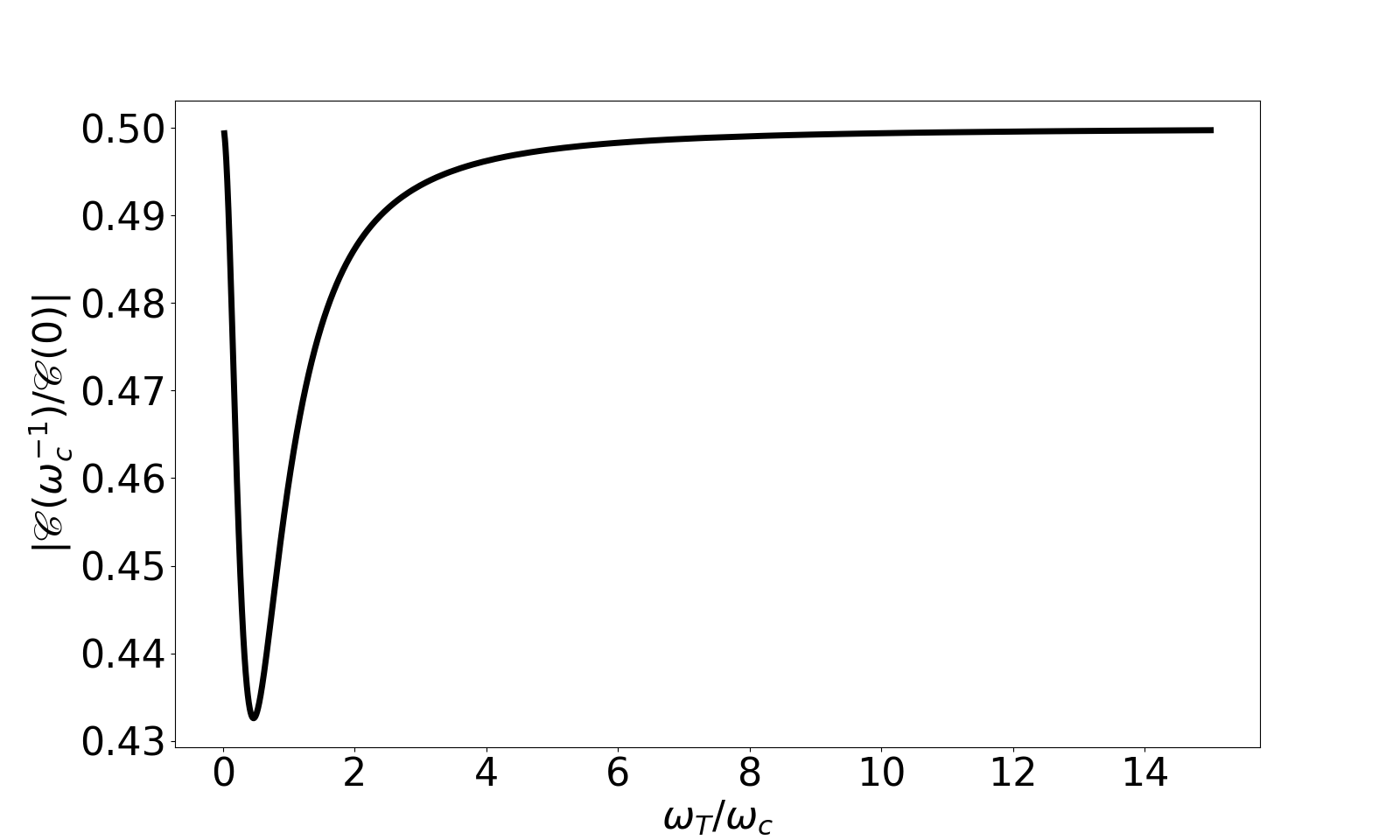}
\caption{$\left|\mathscr{C}\mleft(1/\omega_{c}\mright)/\mathscr{C}\mleft(0\mright)\right|$ as a function of $\omega_{T}/\omega_{c}$, showing that we can use $t_{c}\propto1/\omega_{c}$ as a guide to defining a legitimate boson bath correlation time since $\left|\mathscr{C}\mleft(1/\omega_{c}\mright)/\mathscr{C}\mleft(0\mright)\right|\rightarrow1/2$ for $\omega_{T}\ll\omega_{c}$ and for $\omega_{T}\gg\omega_{c}$. We also see that even when $\omega_{T}\approx\omega_{c}$ we still
can use $t_{c}\approx1/\omega_{c}$ as a good estimate of the bath
correlation time. To ensure a margin of safety to have the correlation function low enough at $t_{c},$ here we define the boson field correlation
time as $t_{c}=2\pi/\omega_{c}$. In this case, $\left|\mathscr{C}\mleft(t_{c}\mright)/\mathscr{C}\mleft(0\mright)\right|\approx0.025.$}
\label{tc figure}
\end{figure}

The two-time correlation function of the boson field, satisfying $\mathcal{C}\mleft(t,t^{\prime}\mright)\equiv\mathscr{C}\mleft(t-t^{\prime}\mright)$,
is given by
\begin{align}
\mathscr{C}\mleft(t\mright) = \mathrm{Tr}_{B}\mleft[B\mleft(t\mright)B\mleft(0\mright)\rho_{B}\mleft(0\mright)\mright],\label{correlation}
\end{align}
where $\mathrm{Tr}_{B}$ is the partial trace over the boson bath
degrees of freedom. It then follows from Eqs. (\ref{B(t)}), (\ref{thermal}),
(\ref{Ohmic}), and (\ref{correlation}) that
\begin{align}
\frac{\mathscr{C}\mleft(t\mright)}{\eta\left(\hbar\omega_{c}\right)^{2}} &= \frac{1}{\left(1+i\omega_{c}t\right)^{2}}+2\left(\frac{\omega_{T}}{\omega_{c}}\right)^{2} \nonumber \\
& \phantom{{}={}} \times\mathrm{Re}\mleft[\psi^{(1)}\mleft(1+\frac{\omega_{T}}{\omega_{c}}\left(1-i\omega_{c}t\right)\mright)\mright],\label{explicit integrals}
\end{align}
where $\omega_{T}=k_{B}T/\hbar$ and $\psi^{(1)}\mleft(z\mright)$
is the first polygamma function~\cite{georgearfken2011}. Equation
(\ref{explicit integrals}) shows that when the temperature is low
enough that $\omega_{T}\ll\omega_{c},$ then $\left|\mathscr{C}\mleft(t\mright)/\mathscr{C}\mleft(0\mright)\right|=1/2$
for $t=1/\omega_{c}.$ When the decoherence is dominated by thermal
noise, namely, when $\omega_{T}\gg\omega_{c},$ either by using the
asymptotic series for the polygamma function~\cite{georgearfken2011}
or by numerical calculation we can see that $\left|\mathscr{C}\mleft(t\mright)/\mathscr{C}\mleft(0\mright)\right|\approx1/2$
for $t\approx1/\omega_{c},$ as shown in Fig.\ref{tc figure}.
Accordingly, for our purposes here, we define $t_{c}=2\pi/\omega_{c}$
as the correlation time of the boson bath, where the $2\pi$ factor
is to give a margin of safety to ensure that the correlation is low
enough at $t_{c}$ for all temperature regimes. Indeed, choosing $t_{c}=2\pi/\omega_{c}$ implies $\left|\mathscr{C}\mleft(t_{c}\mright)/\mathscr{C}\mleft(0\mright)\right|\approx0.025.$

\subsection{Purification Process}\label{qoc}

As exposed, our main objective in this manuscript is to determine the optimal control Hamiltonian that, despite the environmental perturbations, evolves any initial
qubit state $\rho_{IS}\mleft(0\mright)$ to a target state $U_{\tau}\rho_{IS}\mleft(0\mright)U_{\tau}^{\dagger},$
where $U_{\tau}$ is a target single-qubit operator, that is, $U_{\tau}\in\mathrm{SU}\mleft(2\mright).$
By optimal we mean that we have a cost function associated with the
way we choose the control Hamiltonian of Eq. (\ref{Hc(t)}) such as,
for instance, when we want the least use of total energy during the
quantum operation, from $t=0$ to $t=\tau,$ which is specified by
the circuit requirements where the gate is to function and deliver
the intended action on the qubit state. Ideally, the optimal control
would evolve a pure state $\rho_{IS}\mleft(0\mright)$ to the intended
pure state $U_{\tau}\rho_{IS}\mleft(0\mright)U_{\tau}^{\dagger},$ in
which case, for perfect compensation against the noise, we would have
a qubit state $\left|\psi\mleft(0\mright)\right\rangle $ transformed
to $U_{\tau}\left|\psi\mleft(0\mright)\right\rangle ,$ where $\rho_{IS}\mleft(0\mright)=\left|\psi\mleft(0\mright)\right\rangle \left\langle \psi\mleft(0\mright)\right|.$

We notice that the perturbations mix the qubit state and we do not
obtain the desired target evolution, admitting that the control protocol
is not perfect. A possible function to consider is the usual Uhlmann-Jozsa
fidelity measure defined by~\cite{Uhlmann1976,jozsa}
\begin{align}
\mathcal{F}_{UJ}\mleft(\rho,\sigma\mright) = \left[\mathrm{Tr}\mleft(\sqrt{\sqrt{\rho}\sigma\sqrt{\rho}}\mright)\right]^{2}.\label{UJfidelity}
\end{align}
Then the cost function can be taken to be the infidelity defined by
$1-\mathcal{F}_{UJ} \mleft(\rho,\sigma \mright)$, where we take $\sigma=U_{\tau}\rho_{IS}\mleft(0\mright)U_{\tau}^{\dagger},$
the desired output, and $\rho=\rho_{IS}\mleft(\tau\mright),$ obtained
by solving Eq. (\ref{master}).

The procedure to determine the optimal control is to minimize the
infidelity for all possible input states $\rho_{IS}\mleft(0\mright),$
given a target $U_{\tau}.$ Further constraints imposed on the optimization
can be incorporated into the definition of the cost function using
Lagrange multipliers. For instance, if we want to get the least infidelity
using the least amount of energy for the control fields, we can minimize
the infidelity summed to the total work done by the control fields
to get the closest to the desired output.

Despite this robust theoretical approach, quantum optimal control in general is a time-consuming computational
task~\cite{PhysRevLett.114.170501}. One obvious reason why an open quantum system takes longer to optimize than a closed one is apparent
by the fact that in the open system case, given a target $U_{\tau},$
we have to cover the whole set of possible initial states $\rho_{SI}\mleft(0\mright).$
On the contrary, in the closed system case we can consider only the
optimal control Hamiltonian to obtain the evolution of the identity
operator to the target $U_{\tau},$ since in this case the dynamics
are unitary. As we intend to minimize the control field frequencies to minimize control energy, we approximately substitute the dynamics
described by the master equation of Eq. (\ref{master}) for a unitary
problem. To emulate the boson perturbations we introduce the interaction
with a single auxiliary qubit instead, as suggested in Chapter 8 of
Ref.~\cite{michaelnielsen2011}. Here, however, we take into account
the specific time dependence presented in Eqs. (\ref{exact}) and
(\ref{decaying coherence}). Using this device we have an effective purified problem of a finite dimension whose master equation, after
tracing out the auxiliary qubit degrees of freedom is given by
\begin{widetext}
\begin{align}
\frac{\mathrm{d}\rho_{IS}^{(e)}\mleft(t\mright)}{dt} = -\frac{h\mleft(t\mright)}{\hbar^{2}}\int_{0}^{t}\mathrm{d}t^{\prime}\,h\mleft(t^{\prime}\mright)\left[S\mleft(t\mright),\left[S\mleft(t^{\prime}\mright),\rho_{IS}^{\left(e\right)}\mleft(t^{\prime}\mright)\right]\right],\label{effective master}
\end{align}
\end{widetext}
where $\rho_{IS}^{(e)}\mleft(t\mright)$ is
the effective reduced density matrix of the qubit and we define the
function $h\mleft(t\mright)$ as
\begin{align}
h\mleft(t\mright) = \frac{\hbar\dot{\mu}\mleft(t\mright)}{2\sqrt{1-\mu^{2}\mleft(t\mright)}}.\label{h(t)}
\end{align}
To derive Eq. (\ref{effective master}) we use $S\mleft(t\mright)$
as given by Eq. (\ref{S(t)}), $\mu\mleft(t\mright)$ as defined in
Eq. (\ref{decaying coherence}), and instead of the Hamiltonian of
Eq. (\ref{Hnew(t)}), we keep the same $H_{c}\mleft(t\mright)$ and
define an effective Hamiltonian as
\begin{align}
H_{e}\mleft(t\mright) = H\mleft(t\mright)+H_{D}\mleft(t\mright),\label{effective H(t)}
\end{align}
with
\begin{align}
H\mleft(t\mright) = H_{c}\mleft(t\mright)\otimes I\label{H(t)}
\end{align}
and a drift Hamiltonian given by $H_{D}\mleft(t\mright)=-h\mleft(t\mright)\sigma_{z}\otimes\sigma_{z}$
or, using Eq. (\ref{h(t)}), 
\begin{align}\label{drift_hamiltonian}
H_{D}\mleft(t\mright) = -\frac{\hbar\dot{\mu}\mleft(t\mright)}{2\sqrt{1-\mu^{2}\mleft(t\mright)}}\sigma_{z}\otimes\sigma_{z}.
\end{align}
Here $I$ stands for the $2\times2$ identity matrix. In the absence
of control and if in this case $H_{c}\mleft(t\mright)\propto\sigma_{z},$
both Eqs. (\ref{master}) and (\ref{effective master}) give the same
final reduced density matrix for the same initial qubit state, with
the proviso that we have to adopt $\left|\psi_{\text{aux}}\mleft(0\mright)\right\rangle =\left(\left|0\right\rangle +\left|1\right\rangle \right)/\sqrt{2},$
where $\sigma_{z}\left|0\right\rangle =\left|0\right\rangle $ and
$\sigma_{z}\left|1\right\rangle =-\left|1\right\rangle ,$ as the
initial state of the auxiliary qubit. The same statement is not true in general
if we have a time dependent control present. There are two limiting situations in which both Eqs. (\ref{master}) and (\ref{effective master}) give the same answer: when the control Hamiltonian of Eq. (\ref{Hc(t)}) varies in a time
scale long enough as compared to $t_{c}$ so that the approximation $S\mleft(t^{\prime}\mright)  \approx  S\mleft(t\mright)$ is valid in Eqs. (\ref{master}), (\ref{HI(t)}), and (\ref{effective master}), as we show in Appendix~\ref{appA}, and when the gate time $\tau$ is short enough with the correlation time $t_{c}$ long enough that the approximations 
\begin{align}
 B\mleft(t^{\prime}\mright) \approx B\mleft(t\mright), \label{approx B}
\end{align} 
$h\mleft(t^{\prime}\mright)\approx h\mleft(t\mright),$ and $\omega _{c} t \ll 1,$ are valid in Eqs. (\ref{master}) and (\ref{effective master}), as we show in Appendix~\ref{appB}. Thus, since we apply this solution in the context of CDD, i.e., within the long correlation time regime, the approximation holds true and Eqs. (\ref{master}) and (\ref{effective master}) are equivalent.

\subsection{Formulation of the control theory}\label{form}

A time-dependent single-qubit gate can be seen as a quantum circuit
of gradually-changing instantaneous (ideal) single-qubit gates, if
we use a regular partition of the scheduled time interval~\cite{SUZUKI1990319}.
Accordingly, let us use a simple cost function to minimize. We know
that the integral of the square root of the control Hamiltonian bounds
the circuit complexity~\cite{https://doi.org/10.48550/arxiv.quant-ph/0502070,Heller2023,Brown2023}.
We can, however, use the integral over the square of the control Hamiltonian
as our cost function, for the sake of simplicity, since then minimizing
such a cost function will also minimize the conventional one with
the square root of the control Hamiltonian~\cite{montgomery2002tour}.
Our aim is then to minimize our cost function with the implicit assumption
that the time evolution of the identity operator to the target unitary
operator satisfies, at each instant, the Schr\"odinger equation with
the effective total Hamiltonian given by Eq. (\ref{effective H(t)}).
We can simplify the formulation of the control problem by using, at
each instant, a Lagrange multiplier that constrains the instantaneous
unitary operator to be given by the Schr\"odinger equation. In the functional
to minimize, we then use an integral over the constraint equation
summed to the integral of the square of the control Hamiltonian:
\begin{widetext}
\begin{align}
J\mleft(H,U,\Lambda\mright) = \int_{0}^{\tau}\mathrm{d}t\,\mathrm{Tr}\mleft[\frac{1}{2}H\mleft(t\mright)H\mleft(t\mright)+\Lambda\mleft(t\mright)\left(i\hbar\frac{\mathrm{d}U\mleft(t\mright)}{\mathrm{d}t}U^{\dagger}\mleft(t\mright)-H_{e}\mleft(t\mright)\right)\mright],\label{J}
\end{align}
\end{widetext}
where we adopt the convention of using the normalized
trace, that is, $\mathrm{Tr}\mleft(I\mright)=1$ for any square matrix
dimension, $\tau$ is the time interval corresponding to the scheduled
duration of the required gate operation, $\Lambda\mleft(t\mright)$
is the co-state (the matrix containing the instantaneous Lagrange
multipliers), and we now use $H\mleft(t\mright),$ $U\mleft(t\mright),$
and $\Lambda\mleft(t\mright)$ as the functions to be varied independently
to minimize the functional $J\mleft(H,U,\Lambda\mright).$ According
to Pontryagin's Maximum Principle~\cite{PRXQuantum.2.030203}, the
control Hamiltonian $H\mleft(t\mright)$ that minimizes Eq. (\ref{J})
is also the optimal Schr\"odinger equation solution that minimizes the
total energy spent during the interval $\tau$ because it minimizes
the integral of the square of $H\mleft(t\mright).$

We write the factor multiplied by $\Lambda\mleft(t\mright)$ with $\left[\mathrm{d}U\mleft(t\mright)/\mathrm{d}t\right]U^{\dagger}\mleft(t\mright)$
because this quantity is an anti-Hermitian traceless matrix and, therefore, belongs
to the Lie algebra $\mathfrak{su}\mleft(2^{2}\mright)$ and not to the
Lie group $\mathrm{SU}\mleft(2^{2}\mright)$ as does $U\mleft(t\mright),$
so that Eq. (\ref{J}) is mathematically consistent. Because the control
is described by the Hamiltonian of Eq. (\ref{H(t)}), we have a sub-Riemannian
geometry~\cite{montgomery2002tour}, since the control Hamiltonian
does not operate in all directions spanning the whole Lie algebra
$\mathfrak{su}\mleft(2^{2}\mright).$ Actually, in the absence of a control, the Schr\"odinger evolution
is restricted to be generated by the drift Hamiltonian, Eq. (\ref{drift_hamiltonian}),
that does not drive the system out of the one-dimensional subspace
of $\mathfrak{su}\mleft(2^{2}\mright)$ generated by $\sigma_{z}\otimes\sigma_{z}.$
Moreover, because we are restricted to control only the physical qubit
and not the auxiliary one, our control subspace is spanned by the
three-dimensional distribution
\begin{align}
\Delta = \left\{ \sigma_{x}\otimes I,\sigma_{y}\otimes I,\sigma_{z}\otimes I\right\} .\label{Delta}
\end{align}
The effective total Hamiltonian, Eq. (\ref{effective H(t)}), can
only drive the system into the sub-algebra of $\mathfrak{su}\mleft(2^{2}\mright)$
generated by the six-dimensional subspace
\begin{align}
\Gamma = \Delta\cup\left\{ \sigma_{x}\otimes\sigma_{z},\sigma_{y}\otimes\sigma_{z},\sigma_{z}\otimes\sigma_{z}\right\} .\label{Gamma}
\end{align}
The reachable sub-group is then obtained through the exponentiation of the sub-algebra elements 
multiplied by a real number. The subset $\Gamma$
generates a sub-algebra because this generated subspace contained in $\mathfrak{su}\mleft(2^{2}\mright)$ is closed under the commutator operation
between its elements. In other words, the control Hamiltonian does
not belong to the same subspace as the drift Hamiltonian, Eq. (\ref{drift_hamiltonian}),
and this fact complicates our control problem~\cite{PhysRevLett.114.170501}.

The geodesic equations we obtain by functional minimization of Eq.
(\ref{J}) are given by~\cite{SWADDLE20173391,Swaddle-dissertation,Perrier_2020}
\begin{align}
i\hbar\frac{\mathrm{d}U\mleft(t\mright)}{\mathrm{d}t} = H_{e}\mleft(t\mright)U\mleft(t\mright),\label{Schrodinger}
\end{align}
\begin{align}
i\hbar\frac{\mathrm{d}\Lambda\mleft(t\mright)}{\mathrm{d}t} = \left[H_{e}\mleft(t\mright),\Lambda\mleft(t\mright)\right],\label{dLambdadt}
\end{align}
and
\begin{align}
\mathcal{P}\left[\Lambda\mleft(t\mright)\right] = H\mleft(t\mright),\label{PLambda}
\end{align}
where $\mathcal{P}$ is the projector onto the sub-Riemannian subspace
spanned by the distribution $\Delta,$ Eq. (\ref{Delta}). We notice
that the co-state belongs to the sub-algebra spanned by $\Gamma,$
Eq. (\ref{Gamma}). We can join Eqs. (\ref{Schrodinger}), (\ref{dLambdadt}),
and (\ref{PLambda}) into a single equation:
\begin{align}
i\hbar\frac{\mathrm{d}U\mleft(t\mright)}{\mathrm{d}t} &= \left\{ H_{D}\mleft(t\mright)\right. \nonumber \\
 & \phantom{{}={}}+ \left.\mathcal{P}\left[U\mleft(t\mright)\Lambda\mleft(0\mright)U^{\dagger}\mleft(t\mright)\right]\right\} U\mleft(t\mright).\label{unique}
\end{align}

The algorithm we must follow is to try and find $\Lambda\mleft(0\mright)$
that, at $t=\tau,$ gives the desired target unitary operator $U_{\tau}\otimes I$
as the solution $U\mleft(\tau\mright)$ obtained by solving Eq. (\ref{unique}).
The essential difficulty with finding $\Lambda\mleft(0\mright)$ from
Eq. (\ref{unique}) is that we would have to invert the projector
$\mathcal{P},$ which is not uniquely determined: there are infinitely
many possible solutions for $\Lambda\mleft(0\mright)$ giving the same
projection of $U\mleft(t\mright)\Lambda\mleft(0\mright)U^{\dagger}\mleft(t\mright)$
for any given $U\mleft(t\mright).$ Even if we use a Riemannian geometry
with a metric that penalizes the directions orthogonal to the ones
in the distribution $\Delta,$ thus avoiding the projector inversion
problem, we still would not have $\mathrm{d}U\mleft(t\mright)/\mathrm{d}t$ at $\tau,$
being now obliged to face this other indeterminacy. Hence, we choose
to try many initial co-states $\Lambda\mleft(0\mright)$ and locate
the one that renders the solution of Eq. (\ref{unique}), $U\mleft(\tau\mright),$
that is closest to the required target $U_{\tau}\otimes I.$ As a
distance between two unitary operators here we adopt the infidelity
measure, given by the fidelity, as defined in Ref.~\cite{Dong2016},
subtracted from the real unit:
\begin{align}
\mathcal{I}\mleft(U_{1},U_{2}\mright) = 1 -\left|\mathrm{Tr}\mleft(U_{1}^{\dagger}U_{2}\mright)\right|,\label{infidelity}
\end{align}
where we use the normalized trace.

\subsection{Optimal Hamiltonian}\label{subsec:Procedure}

Our aim is to be able to efficiently obtain the co-state $\Lambda\mleft(0\mright)$
once we are given a target $U_{\tau}\otimes I.$ However, the algorithm
we described is based on searching for the minimum of $\mathcal{I}\mleft(U_{\tau}\otimes I,U\mleft(\tau\mright)\mright),$
Eq. (\ref{infidelity}), viewed as a noninvertible map that takes
$\Lambda\mleft(0\mright)$ to $U\mleft(\tau\mright)$ through Eq. (\ref{unique}). Given a target, this search can take a relatively long
time, depending on the required level of accuracy. Here, we show how to compute $\Lambda\mleft(0\mright)$ using a brute-force approach, employing a computationally demanding process named $q\text{-jumping}$ \cite{PhysRevLett.114.170501}. In Sec. \ref{nn}, after calculating a sufficient number of cases using the $q\text{-jumping}$ strategy, we then train a neural network to predict $\Lambda\mleft(0\mright)$ for each given
target $U_{\tau}\otimes I$. As we will show, these predictions can
give estimates that are not only quickly calculated, but also much closer to the desired answer.

The initial guess for $\Lambda\mleft(0\mright)$ is important since
it helps to find one that is the closest to the one that minimizes
the infidelity, Eq. (\ref{infidelity}). If we could invert the projector
operator in $\mathcal{P}\left[U\mleft(t\mright)\Lambda\mleft(0\mright)U^{\dagger}\mleft(t\mright)\right]$
in Eq. (\ref{unique}), then we could easily find an initial guess
for the searching algorithm. The idea would be to write
$U_{\tau}=\exp\mleft(-iH_{\text{guess}}\tau/\hbar\mright),$ then we would
calculate $H_{\text{guess}}$ and use $H_{\text{guess}}=H_{D}\mleft(0\mright)+\mathcal{P}\left[\Lambda\mleft(0\mright)\right]$
to find an initial guess for $\Lambda\mleft(0\mright),$ but we can
not invert $\mathcal{P}.$ To avoid this problem we proceed in an
alternative way. We use a Riemannian geometry with penalty factors
along the directions in $\Gamma,$ Eq. (\ref{Gamma}), that do not
belong to the distribution $\Delta,$ Eq. (\ref{Delta}). Inspired
by the methodology of Ref.~\cite{PhysRevLett.114.170501}, our formulation follows.

For our purposes, we use a single penalty factor, $q,$ along the directions
in the difference set:
\begin{align}
\Gamma\backslash\Delta = \left\{ \sigma_{x}\otimes\sigma_{z},\sigma_{y}\otimes\sigma_{z},\sigma_{z}\otimes\sigma_{z}\right\} .\label{difference}
\end{align}
We now define a control Hamiltonian that includes all six directions
of $\Gamma:$
\begin{align}
H_{q}\mleft(t\mright) = \sum_{k=1}^{3}\alpha_{k}\mleft(t\mright)\sigma_{k}\otimes I+\sum_{k=1}^{3}\beta_{k}\mleft(t\mright)\sigma_{k}\otimes\sigma_{z},\label{Hq}
\end{align}
where $\left(\sigma_{1},\sigma_{2},\sigma_{3}\right)=\left(\sigma_{x},\sigma_{y},\sigma_{z}\right)$
and $\alpha_{k}\mleft(t\mright),\beta_{k}\mleft(t\mright)$ for $k=1,2,3,$
are real functions of time. Now we proceed analogously as when we
minimized Eq. (\ref{J}), except that now, instead of the square of
the control Hamiltonian in the integrand, we use $G_{q}\mleft(H_{q}\mleft(t\mright),H_{q}\mleft(t\mright)\mright),$
where, given two Hamiltonians, $H_{A}$ and $H_{B},$ we define the
Riemannian metric
\begin{align}
G_{q}\mleft(H_{A},H_{B}\mright) &= \frac{1}{2}\sum_{k=1}^{3}\mathrm{Tr}\mleft[H_{B}\left(\sigma_{k}\otimes I\right)H_{A}\mright] \nonumber \\
 & \phantom{{}={}} + \frac{q}{2}\sum_{k=1}^{3}\mathrm{Tr}\mleft[H_{B}\left(\sigma_{k}\otimes\sigma_{z}\right)H_{A}\mright],\label{Gq}
\end{align}
where $q$ is a finite positive real number. By increasing $q,$ the
equations that give the geodesic must converge to the desired case
of our sub-Riemannian metric where the control is in the form of Eq.
(\ref{Hc(t)}), since the solution to the minimization problem gives
$\beta_{k}\mleft(t\mright)\rightarrow0,$ for $k=1,2,3,$ as $q\rightarrow\infty.$

The new geodesic equation obtained for a finite $q$ is given by
\begin{align}
i\hbar\frac{\mathrm{d}U_{q}\mleft(t\mright)}{\mathrm{d}t} &= \left\{ H_{D}\mleft(t\mright)\right.\nonumber \\ &\phantom{{}={}} + \left.\mathcal{F}_{q}\left[U_{q}\mleft(t\mright)\Lambda_{q}\mleft(0\mright)U_{q}^{\dagger}\mleft(t\mright)\right]\right\} U_{q}\mleft(t\mright),\label{q-Schrodinger}
\end{align}
with a new invertible operator, for finite $q,$ defined by
\begin{align}
\mathcal{F}_{q} = \mathcal{P}+\frac{1}{q}\mathcal{Q},\label{Fq}
\end{align}
where $\mathcal{P}$ is the same projector of Eq. (\ref{PLambda}),
that projects onto the subspace spanned by the distribution $\Delta,$
while $\mathcal{Q}$ is the projector onto the subspace spanned by
$\Gamma\backslash\Delta,$ Eq. (\ref{difference}). The operator $\mathcal{F}_{q}$
is not a projector for finite $q,$ but $\mathcal{F}_{q}\rightarrow\mathcal{P}$
as $q\rightarrow\infty.$

We start with $q=10$ and, for each target $U_{\tau}$ generated according
to Eq. (\ref{Utau}), we write $U_{\tau}=\exp\mleft(-iH_{\text{guess}}\tau/\hbar\mright)$
and calculate $H_{\text{guess}}.$ Then we impose $H_{\text{guess}}=H_{D}\mleft(0\mright)+\mathcal{F}_{q=10}\left[\Lambda_{q=10}\mleft(0\mright)\right]$
to find an initial guess for $\Lambda_{q=10}\mleft(0\mright),$ since
we can invert $\mathcal{F}_{q=10}.$ Numerically, we can use, for
instance, \texttt{FindMinimum} in Wolfram Language or \texttt{scipy.optimize.minimize}
in Python to discover, from our initial guess, a better $\Lambda_{q=10}\mleft(0\mright)$
that minimizes the infidelity, Eq. (\ref{infidelity}). The first
trial sometimes takes a relatively long time, depending on the target
$U_{\tau},$ and in some cases the infidelity is hardly as low as we require.
We then refine the result by using a shooting algorithm based on the
Newton method to find the value of $\Lambda_{q=10}\mleft(0\mright),$
starting from the output of \texttt{FindMinimum} or \texttt{scipy.optimize.minimize},
that is a zero of the infidelity.

We then iterate the process named $q\text{-jumping}$ in Ref.~\cite{PhysRevLett.114.170501}
by using the refined $\Lambda_{q=10}\mleft(0\mright)$ as an initial
guess for the evolution for the case with $q>10.$ The iteration process
proceeds by increasing $q$ by an increment $\Delta q$ and using
$\Lambda_{q}\mleft(0\mright)$ obtained from the previous iteration
as a first guess for the evolution with a new penalty factor of $q+\Delta q.$
After numerically minimizing, if the infidelity is not low enough
(we could pick a tolerance on the order of $10^{-5},$ for instance),
we use the shooting algorithm to improve the result for $\Lambda_{q+\Delta q}\mleft(0\mright).$
Sometimes, however, depending on the target $U_{\tau},$ the minimization
is not near enough to the minimum of the infidelity that the shooting
method diverges even further from the minimum of Eq. (\ref{infidelity}).
In such cases, we keep the $\Lambda_{q+\Delta q}\mleft(0\mright)$ that
we had as output of the numerical minimization.

The choice of $\Delta q$ for each iteration is multiplicative and
it is not fixed. Because after a gradual increase to a certain value
of $q$ the difference between $\mathcal{F}_{q}$ and $\mathcal{P}$
becomes small enough, the jump from a large enough $q$ to $q\rightarrow\infty$
becomes easily implemented. However, for intermediary values of $q,$
typically lower than $100,$ the jumps have to be smaller than for
$q>100$ between one iteration and the next. We, therefore, by trial
and error, define a way to increase $q$ in a multiplicative way so
that after a chosen number of iterations, $N_{\text{it}},$ we reach a maximum
$q=q_{\text{max}}$ value of at least $2000$ or more (up to $10000$ in
some easier cases), using a multiplicative factor $10^{\chi}$ with
$\chi=\log_{10}\left(q_{\text{max}}/q_{\text{in}}\right)/N_{\text{it}},$ where $q_{\text{in}}$
is the initial $q$, which in the present description we picked as $q_{\text{in}}=10.$
If, depending on the target $U_{\tau}$, the required tolerance is
not reached, we use a decreasing multiplicative factor and retake
the iteration step so that we try and see if with a lower increase
of $q$ the guess is good enough to converge to a lower infidelity
than the infidelity with the previous intended $q$ increment. Because of this some
easier cases reach up to, say, $q=10000$ with good fidelity after
$N_{\text{it}},$ while other, more difficult cases, do not even reach, with
low enough infidelity, the minimum desired $q=2000.$ For  difficult
cases, we then run another pass starting over from the beginning,
with $q=10,$ with even lower increments of $q$ between iterations. After determining $\Lambda_{q_{\text{max}}}\mleft(0\mright)$
with high enough fidelity, we run the sub-Riemannian dynamics to take
the $q\text{-jump}$ to infinity using the evolution given by Eq.
(\ref{unique}) with $\Lambda_{q_{\text{max}}}\mleft(0\mright)$ as initial
guess. Thus, after this quite demanding computational effort, we determine $\Lambda\mleft(0\mright)$, which provides the optimal field control for any quantum gate. This result marks an advancement in the CDD strategy, since once the environment is well characterized, an optimal control field can be determined with high precision.

\subsection{Control fields for amplitude damping}\label{subsec:ampdamp}
The described method for finding the optimal Hamiltonian was done for the specific case where we use the effective Hamiltonian given by Eq.~(\ref{drift_hamiltonian}), which causes dephasing. Since the process is computationally demanding, it is pertinent to ask how the control fields must change if we consider another type of noise. Amplitude damping as an example.

With some considerations, it can be shown that a slight modification of the control fields for dephasing can be used for protection against amplitude damping without the need to perform the q-jumping from scratch. To see that, analogously to Eq.~(\ref{Htot(t)}), let us model the amplitude damping with a usual Jaynes-Cummings interaction as
\begin{align}
    H_{\text{tot}}\mleft(t\mright) &= H_{c}\mleft(t\mright) + \hbar \omega_0 \sigma_z + \sum_{\lambda}\hbar\omega_{\lambda}b_{\lambda}^{\dagger}b_{\lambda} \nonumber \\
 &\phantom{{}={}} + \sum_{\lambda}\hbar\left(g_{\lambda} \sigma_+ b_{\lambda} + g_{\lambda}^{\ast} \sigma_- b_{\lambda}^{\dagger}\right),
\end{align}
where the control for this case is given by $H_c(t) = \hbar f_x(t) \sigma_x + \hbar f_y(t) \sigma_y + \hbar f_z(t) \sigma_z$ and $\hbar \omega_0$ is the energy gap between the two levels. If we assume the rotating-wave approximation (RWA) as valid, it means that the energy non-conserving terms proportional to $g_\lambda^*\sigma_+ b_\lambda^\dagger$ and $g_\lambda \sigma_- b_\lambda$ are negligible, so they can be added to the Hamiltonian without much impact, yielding
\begin{align}\label{CRterms}
    H_{\text{tot}}\mleft(t\mright) &\approx H_{c}\mleft(t\mright) + \hbar \omega_0 \sigma_z + \sum_{\lambda}\hbar\omega_{\lambda}b_{\lambda}^{\dagger}b_{\lambda} \nonumber \\
 &\phantom{{}={}} + \sum_{\lambda}\hbar\left(g_{\lambda} \sigma_+ b_{\lambda} + g_{\lambda}^{\ast} \sigma_- b_{\lambda}^{\dagger}\right) \nonumber \\
 &\phantom{{}={}} + \sum_{\lambda}\hbar\left(g_{\lambda}^* \sigma_+ b_{\lambda}^\dagger + g_{\lambda} \sigma_- b_{\lambda}\right),
\end{align}
Then, by doing the unitary transformation given by $U_{B}\mleft(t\mright)=\exp\mleft(-it\sum_{\lambda}\omega_{\lambda}b_{\lambda}^{\dagger}b_{\lambda}\mright)$ and considering $\sigma_\pm = \frac{1}{2}\left( \sigma_x \pm i \sigma_y \right)$ we obtain the new Hamiltonian
\begin{align}
    H_\text{new}(t) = H_c(t) + \hbar \omega_0 \sigma_z + \sigma_x B(t).
\end{align}
By identifying the first two terms as an effective control $H_c^\text{eff}(t) \equiv H_c(t) + \hbar \omega_0 \sigma_z$ the equation has the exact same form as Eq.~(\ref{Hnew(t)}) with the only difference being the operator $\sigma_x$ instead of $\sigma_z$. So if we do a cyclic permutation of the axes as $\{x, y, z\} \rightarrow \{z, x, y\}$ and appropriately identify the components of $H_c^\text{eff}(t)$ with those of $H_c(t)$ obtained with the q-jumping for the dephasing noise, the equation should be the same and $H_c^\text{eff}(t)$ should be able to protect the qubit against amplitude damping noise. Explicitly, this means that given the control fields $\omega_x(t)$, $\omega_y(t)$ and $\omega_z(t)$, calculated with the q-jumping method for dephasing noise, the control fields for protecting against amplitude damping noise should follow the relation
\begin{align}
    f_x(t) &= \omega_z(t), \label{fx} \\
    f_y(t) &= \omega_x(t), \label{fy} \\
    f_z(t) &= \omega_y(t) - \omega_0. \label{fz} 
\end{align}

As was mentioned, these control fields will only be suitable in a situation where the RWA is valid. To determine the condition for it we may first write the total Hamiltonian in the absence of a control field and in the representation without the environment Hamiltonian, which reads
\begin{align}
    H(t) &= \hbar \omega_0 \sigma_z \nonumber \\
    &\phantom{{}={}} + \sum_\lambda \hbar \left( g_\lambda e^{-i \omega_\lambda t} \sigma_+ b_\lambda + g_\lambda^* e^{i \omega_\lambda t} \sigma_- b_\lambda^\dagger \right) \nonumber \\
    &\phantom{{}={}} + \sum_\lambda \hbar \left( g_\lambda^* e^{i \omega_\lambda t} \sigma_+ b_\lambda^\dagger + g_\lambda e^{-i \omega_\lambda t} \sigma_- b_\lambda \right),
\end{align}
and writing it in the interaction picture will yield
\begin{align}
    H_I(t) &= \sum_\lambda \hbar \left( g_\lambda e^{-i (\omega_\lambda-\omega_0) t} \sigma_+ b_\lambda \right. \nonumber \\
    &\phantom{{}=\sum_\lambda \hbar \left(\right.} + \left. g_\lambda^* e^{i (\omega_\lambda-\omega_0) t} \sigma_- b_\lambda^\dagger \right) \nonumber \\
    &\phantom{{}={}} + \sum_\lambda \hbar \left( g_\lambda^* e^{i (\omega_\lambda+\omega_0) t} \sigma_+ b_\lambda^\dagger \right. \nonumber \\
    &\phantom{{}= + \sum_\lambda \hbar \left(\right.} + \left. g_\lambda e^{-i (\omega_\lambda+\omega_0) t} \sigma_- b_\lambda \right).
\end{align}
When we consider that the coupling constants follow the Ohmic distribution, given by Eq.~(\ref{Ohmic}), we have a peak at $\omega = \omega_c$. For the RWA to be valid, we need to be able to neglect the terms proportional to $e^{\pm i (\omega_\lambda + \omega_0) t}$ when compared to the ones proportional to $e^{\pm i (\omega_\lambda - \omega_0) t}$, and that is only possible if $\omega_0 \approx \omega_c$. If this condition is not met then the approximation done in Eq.~(\ref{CRterms}) is not justified and the fields given by Eqs.~(\ref{fx})-(\ref{fz}) will not be suitable for protection against amplitude damping.

Another important point is that the fields do not only protect against external interactions. The desired quantum gate is also applied, and since the entire configuration of fields will be rotated, the applied gate will not, in general, correspond to the desired one. Specifically, if we apply the fields for the $X$ gate, the result will be a protected $Y$ gate. The reason is that, with the permuted axes, the X axis now corresponds to the second coordinate, which is Y. The same happens for the other two. If we apply the fields of the $Y$ gate, the result will be a $Z$ gate, and if we apply the fields of $Z$, the result will be an $X$ gate.

To address this problem, consider we desire to apply a general gate written as
\begin{align}\label{desired_gate}
    U = a \mathbb{I} + i \left( bX + cY + dZ \right),
\end{align}
with $a,b,c,d \in \mathbb{R}$ and $a^2+b^2+c^2+d^2=1$. 
If we calculate the optimal control for protection against dephasing noise and simply use the rotated fields for the amplitude damping case, the result will be a protected gate 
\begin{align}
    \widetilde{U} = a \mathbb{I} + i \left( bY + cZ + dX \right).
\end{align}
Now consider the following unitary operators:
\begin{align}
    C_1 &\equiv \frac{-X+Z}{\sqrt{2}}, \\
    C_2 &\equiv \frac{Y-Z}{\sqrt{2}}.
\end{align}
It can be verified that, defining a ``correction'' operator $U_\text{cor} \equiv C_2 C_1$, we have
\begin{align}
    U_\text{cor}^\dagger X U_\text{cor} &= Z, \\
    U_\text{cor}^\dagger Y U_\text{cor} &= X, \\
    U_\text{cor}^\dagger Z U_\text{cor} &= Y.
\end{align}
Therefore, finding the optimal control of the gate
\begin{align}
    U_\text{cor}^\dagger U U_\text{cor} = a \mathbb{I} + i \left( bZ + cX + dY \right)
\end{align}
in the dephasing case will yield the desired gate $U$ from Eq.~(\ref{desired_gate}) in the amplitude damping case.

This strategy can be extended to any interaction of the form
\begin{align}\label{general_interaction}
    H_\text{int} = \boldsymbol{\sigma} B(t),
\end{align}
where $\boldsymbol{\sigma} = a \sigma_x + b\sigma_y + c\sigma_z$ with $a,b,c \in \mathbb{R}$ and $a^2 + b^2 + c^2 = 1$, given that we calculate the optimal control for the appropriate ``correction'' gate. In general, for an interaction of the form shown in Eq.~(\ref{general_interaction}), a direct application of gates $X, Y$, and $Z$ would result in arbitrary gates $U_X \equiv R^\dagger X R$, $U_Y \equiv R^\dagger Y R$, and $U_Z\equiv R^\dagger Z R$, and the correction operator would be the one such that $U_\mathrm{cor} = R^\dagger$. And then, if one desired to apply any gate $U$ in the case of a noise that can be described by the interaction in Eq.~(\ref{general_interaction}), they would need to calculate the optimal control for the gate $U_\text{cor}^\dagger U U_\text{cor}$ and use the fields with the appropriately rotated components. The result would be a protected $U$ gate.

\section{Neural Network}\label{nn}
As we mentioned above, the computational effort needed to compute a single gate is considerable. This becomes particularly problematic in the context of circuit optimization, where on-demand logical gates are essential. To treat this problem, we develop a neural network that can learn from previous results. 

We begin generating an initial population of $2000$ quantum gates using
\begin{align}
U_{\tau} = I\cos{\theta}-i\mathbf{\hat{u}}\boldsymbol{\cdot}\boldsymbol{\sigma}\sin{\theta},\label{Utau}
\end{align}
where the values of the versor $\mathbf{\hat{u}}$ are randomly chosen
to cover the whole unit sphere and the values of $\theta$,
also randomly picked, cover the upper half of the unit circumference.
Then, instead of using $U_{\tau}$, we store only the values of $\theta\mathbf{\hat{u}}$,
since we can easily recover the target unitary operator by using $U_{\tau}=\exp\mleft(-i\theta\mathbf{\hat{u}}\boldsymbol{\cdot}\boldsymbol{\sigma}\mright).$

Because $\Lambda\mleft(0\mright)$ is a Hermitian matrix given as a
linear combination of the elements in $\Gamma,$ Eq. (\ref{Gamma}),
we only store the six corresponding real coefficients. Our aim is
then to find these six real quantities once we are given the three
real values $\theta\mathbf{\hat{u}}$ defining the target $U_{\tau}.$
Naturally, we train and test this neural
network to, afterward, guess target cases not in the initial training
set of nearly $ 2000$  cases (some cases could not reach high enough fidelity).
We then check the infidelity of the guessed $\Lambda\mleft(0\mright)$
using it as an initial value in Eq. (\ref{unique}) and, when necessary,
we use numerical minimization as before to refine the result to a
tolerable infidelity. Now, however, it turns out that the process
of guessing, deciding if refinement is necessary, and eventual refinement
is much faster than the cumbersome $q\text{-jumping}$ algorithm we
described. In the next subsections, we present the results of our investigations.

\subsection{Simple Example}

To illustrate all the steps of the procedure we described in the previous
section, let us consider a single case. In the following, we use the
target unitary operator given by Eq. (\ref{Utau}), namely,
\begin{align*}
U_{\tau} = \exp\mleft(-i\theta\mathbf{\hat{u}}\boldsymbol{\cdot}\boldsymbol{\sigma}\mright),
\end{align*}
with
\begin{align}
\theta\mathbf{\hat{u}} = 0.307485\mathbf{\hat{x}}+0.346931\mathbf{\hat{y}}-2.78627\mathbf{\hat{z}}.
\label{Reginaldo_unitary}
\end{align}
In the code, the complete $U_{\tau}\otimes I$ unitary operator is
represented only by the coefficients $u_{k}$ in $(0.307485,0.346931,-2.78627,0,0,0),$
since using the  elements $\gamma_{k}$ of $\Gamma,$ Eq. (\ref{Gamma}),
namely, $\sigma_{k}\otimes I,$ for $k=1,2,3,$ and $\sigma_{k-3}\otimes\sigma_{z},$
for $k=4,5,6,$ we have
\begin{align*}
U_{\tau}\otimes I = \exp\mleft(-i\sum_{k=1}^{6}u_{k}\gamma_{k}\mright).
\end{align*}

According to Sec. \ref{subsec:Procedure}, we next change the previous
code vector to $(0.307485, 0.346931, -2.78627, 0, 0, q h\mleft(0\mright)),$
now storing the coefficients for $\left[H_{\text{guess}}-q H_{D}\mleft(0\mright)\right],$
and we use these components as coefficients of the $\Gamma$ basis
matrices to define our initial guess for $\Lambda_{q=10}\mleft(0\mright).$

Our code took about forty minutes to run $100$ jumps, on one of our
desktop computers, and we ended up propagating $q$ up to the value
$q = 41.9348$, reaching infidelity of $0.0000999889$, and giving
a $\Lambda_{q}\mleft(0\mright)$ stored in code with coefficients 
%(1.11881,0.891343,0.277546,-1.84379,1.09701,-19.2034).
\begin{align*}
    (1.1188,0.89134,0.27755,-1.8438,1.0970,-19.2034).
\end{align*}
For this calculation, we used a maximum infidelity tolerance of $10^{-4}$.

We then can take the result of the previous calculation and restart
the $q-$jumping, now relaxing the tolerance to, say, $5\times10^{-3}$.
After other $100$ jumps we ended up with $q=2000,$ reaching an infidelity
of $0.000728164$, and obtaining a $\Lambda_{q}\mleft(0\mright)$ stored
in code as 
%$(9.74364,0.542157,0.57394,5.52942,-22.2162,-39.9236).$
\begin{align*}
    (9.7436,0.54216,0.5739,5.5294,-22.2162,-39.9236).
\end{align*}
This calculation took about $57$ minutes on the same desktop computer.

We can now continue the $q\text{-jumping}$ process using the same
tolerance of infidelity, up to $5\times10^{-3},$ taking now $200$
jumps from $q=2000$ to $q=10000$. This run took about $76$ minutes,
reaching an infidelity of $0.000661113$, and resulting in a $\Lambda_{q}\mleft(0\mright)$
stored as 
%$(10.5295,0.382698,0.557801,5.30813,-24.297,-44.0305).$
\begin{align*}
    (10.5295,0.3827,0.5578,5.3081,-24.2970,-44.0305).
\end{align*}
From this result now we run the jump for the limit in which $q$ tends to an infinite value, obtaining 
%$(10.5288,0.38277,0.557781,5.30812,-24.297,-44.0307),$
\begin{align*}
    (10.5288,0.3828,0.5578,5.3081,-24.2970,-44.0307),
\end{align*}
with an infidelity of $0.000674937.$ The jump to $q \rightarrow \infty$ is
much faster: this one took only $22$ seconds.

\subsection{Neural network predictions}

To make use of neural networks, as inspired by Refs. \cite{SWADDLE20173391,Perrier_2020}, we
built many models searching for a good architecture. In contrast to both authors, we did
not make use of recurrent neural networks. Instead, we built and optimized a simple fully connected perceptron.
\cite{chollet_deep_2018, Goodfellow-et-al-2016}
We want to predict the relationship between 
\(\lambda_k\), such that \(\Lambda(0) = \sum_{k=1}^6 \lambda_k \gamma_k\), 
and each of the desired \(u_k\). Therefore, we designed the network to use \(u_k\) as
its input, and predict the components \(\lambda_k\). The loss used in training is the
usual MSE.

We trained a neural network model using the population of
\(2000\) targets, using it to predict the costate \(\Lambda(0)\)
that generates the vector shown in Eq. \eqref{Reginaldo_unitary}. The model itself 
took around \(0.1\) second to obtain a prediction for the costate, and the infidelity 
of said prediction is around \(0.33\). However, with this prediction we can use 
\texttt{scipy.optimize.minimize} in order to refine \(\Lambda(0)\), taking about \(90\)
seconds to obtain an infidelity of \(\approx 9.2\times10^{-11}\). Notice that we can use the model to
predict in seconds new unitaries that would take hours by using the \(q\)-jumping method. 

With the \(2000\) targets, we can start a process of data augmentation by using the 
said network itself. To do so, we generate \(100\) new unitary targets, find initial guesses
with the network, process them through the minimization algorithm, and add them to the 
dataset. This dataset is used to train the network once again, using only the data that
obtained infidelity smaller than the chosen threshold. This process was repeated 
until our model had around \(6000\) data points available for training. With the new data, we 
trained yet another model, separating around a third of the data as test-only.

With all the new data, we looked for a new model that fits nicely to our data, predicting small infidelities for the desired unitaries. In particular, the specific model has \(5\) hidden layers, 
all using a rectified linear function as their activation function, as well as a dropout 
rate of \(30\%\) and an \texttt{l2} parameter of \(0.002\).

\begin{figure}
    \includegraphics[width = 0.45\textwidth]{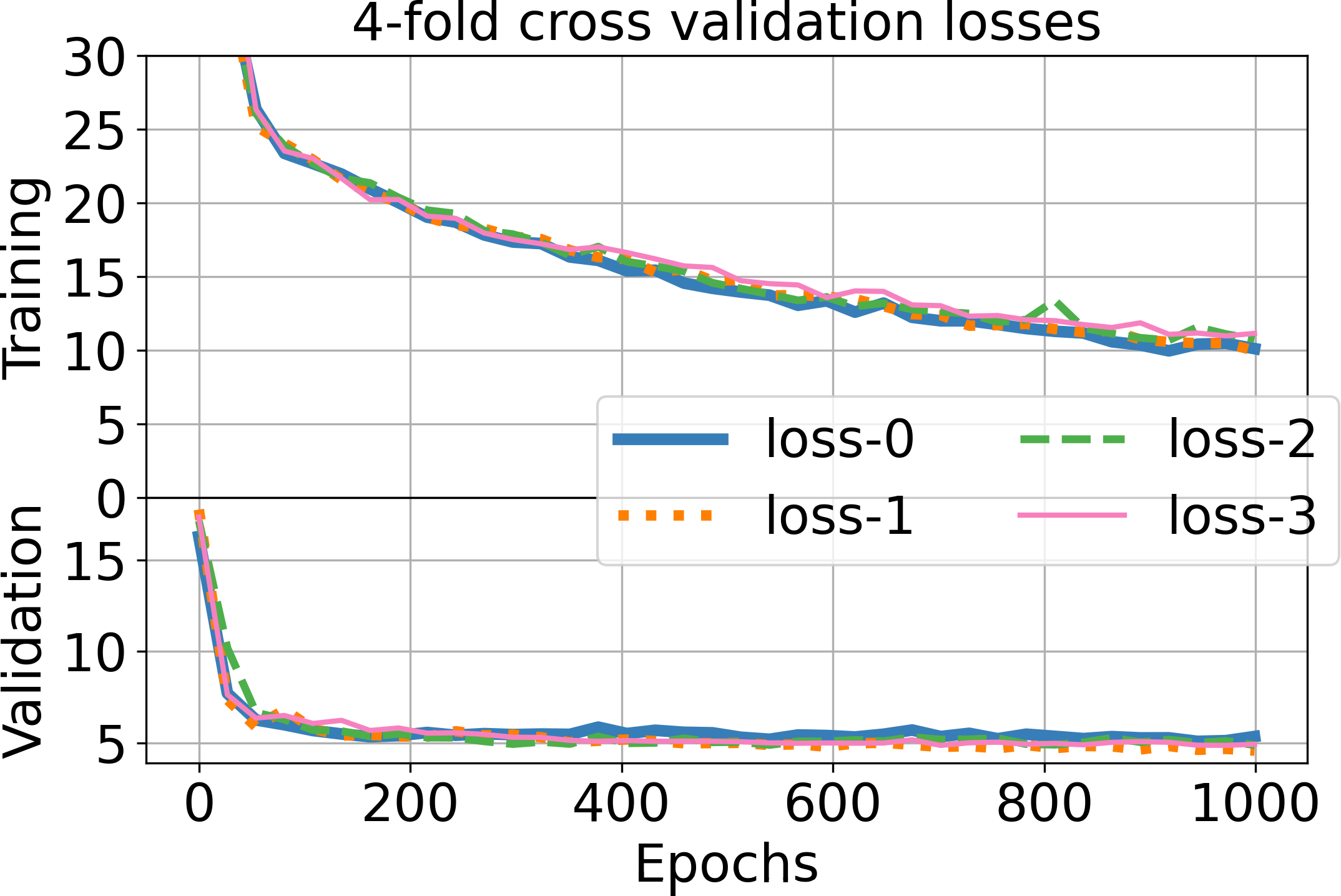}
    \caption{Training and validation losses for the neural network training process.
    With a small dataset, we decided to make use of the \(k\)-fold cross-validation technique for \(k=4\).
    We notice that the validation loss does not decrease after \(200\) epochs.}
    \label{fig:nicolas:kcrossval}
\end{figure}

Afraid of obtaining biased results due to few data - even with the augmentation -, we 
trained the network by applying a k-fold cross-validation test. \cite{chollet_deep_2018, Goodfellow-et-al-2016} The loss result is shown in Fig.
\ref{fig:nicolas:kcrossval}. We notice a plateau in the validation loss around \(200\) epochs. The chosen loss is the usual MSE.

\begin{figure}
    \includegraphics[width = 0.45\textwidth]{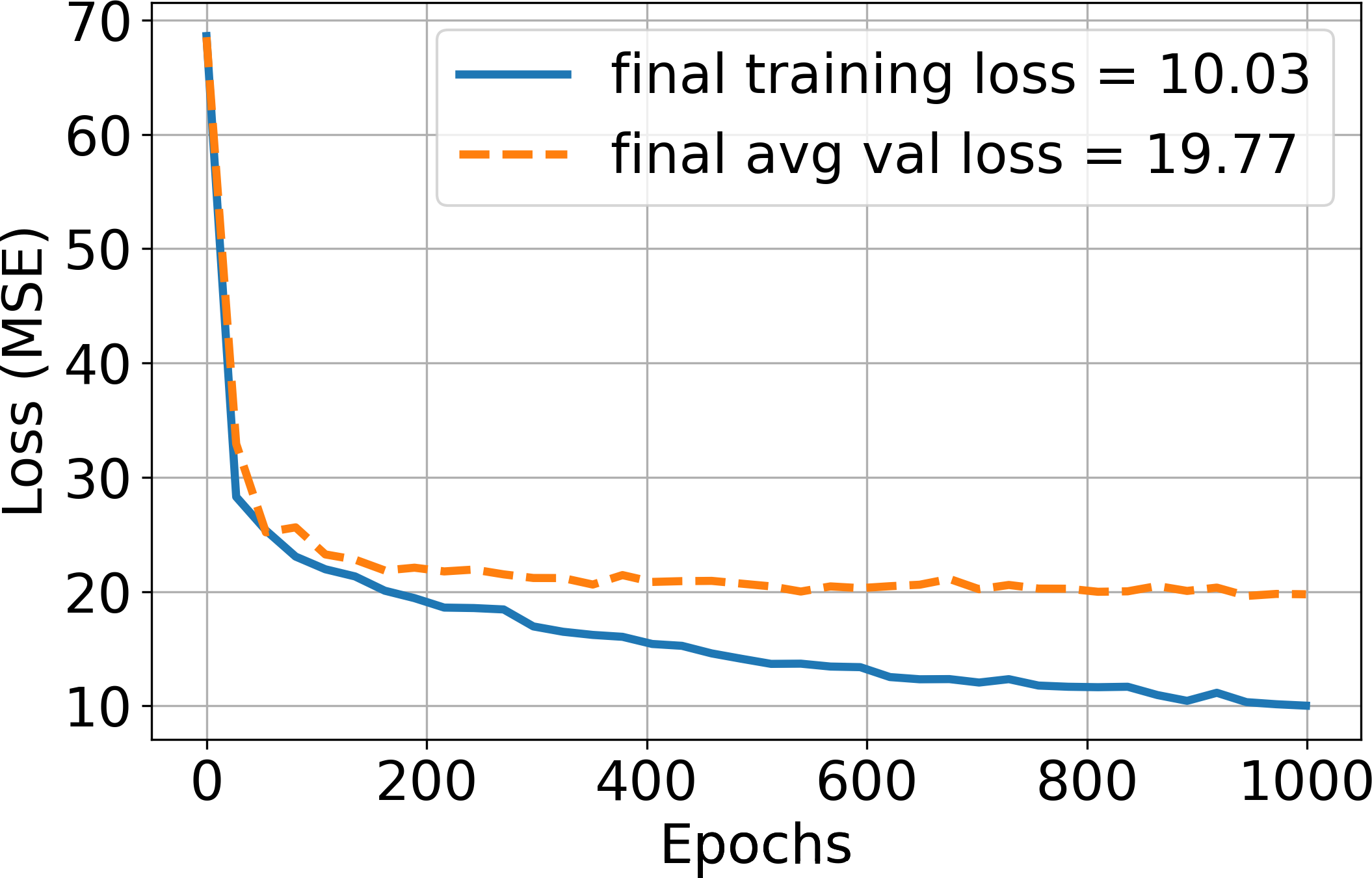}
    \caption{Training loss (continuous line) using all available data on training, compared to the average validation loss (dashed line) obtained
    in the \(k\)-cross validation shown previously.}
    \label{fig:nicolas:all_data_loss}
\end{figure}

Then, we trained the same model on all available data, with the loss presented in Fig. \ref{fig:nicolas:all_data_loss}. 
This plot shows the training loss continuously decreasing in all training, while the validation loss gets stagnant around 
\(200\) epochs as said previously.

\begin{figure}
    \includegraphics[width = 0.47\textwidth]{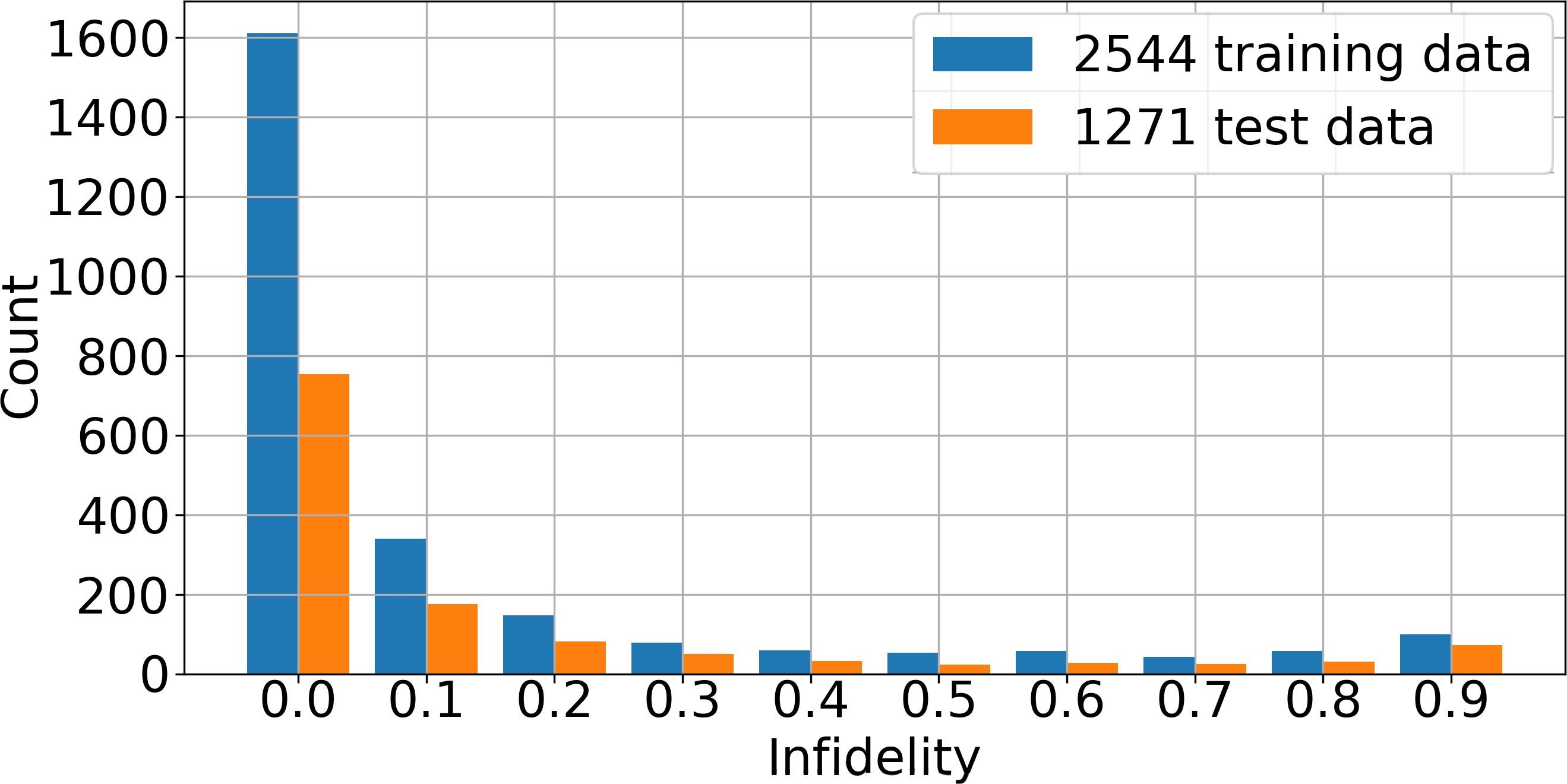}
    \caption{Histogram of infidelities. The first bin - which amounts to around 63\%/59\% of the training/test data -
    comprehends infidelities from \(10^{-4}\) up to \(10^{-1}\). This indicates the neural network predictions are, in general, great guesses to use other numerical minimization tools.}
    \label{fig:nicolas:infidelity_test}
\end{figure}

With this neural network model, we compute the infidelities the unitaries predicted both using the training dataset, as well as the 
test dataset - data that the network has never seen before, not even in validation steps. 
Such infidelities are presented in Fig. \ref{fig:nicolas:infidelity_test}, reaching as low as \(10^{-4}\).
About 63\% of the data is present on the smallest infidelity bin for the training dataset and 59\% for the test dataset.
This means that the output from the network, in many cases, does not need to be significantly improved, although the optimization of such a guess is fast.

\begin{figure}
    \includegraphics[width = 0.465\textwidth]{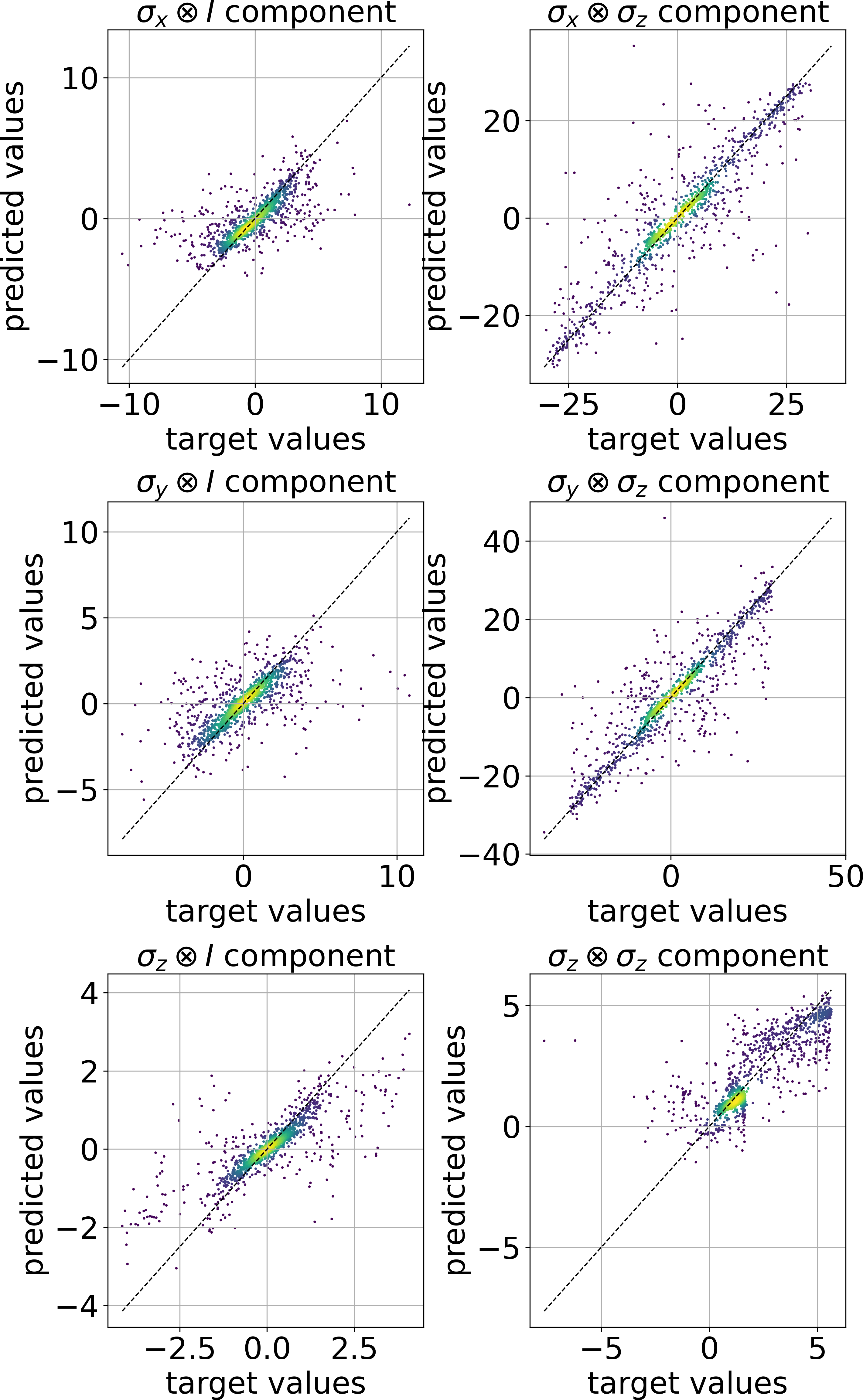}
    \caption{Scatter plot in which we compare the predicted values with the true values for each component
    \(\lambda_k\), using the test dataset. We notice that most components follow the identity curve with little variance. Notice that the off-control components reach higher values than the control ones, taking into account the existence of the drift Hamiltonian. The \(\lambda_6\) components are skewed, probably due to the drift Hamiltonian having the form \(H_D(t) = -h(t)\sigma_z\otimes\sigma_z\). Color indicates density.}
    \label{fig:nicolas:true_vs_true}
\end{figure}

In Fig. \ref{fig:nicolas:true_vs_true}, we show the relation Actual vs. Prediction for \(\lambda_{\text{target}}\) and \(\lambda_{\text{pred}}\) for the test dataset.
Notice that the prediction is accurate to the target values as most of the points are close to the identity curve. In particular, we notice that the \(\lambda_6\) components are skewed. This happens because of the existence of the external Hamiltonian, which drifts the system towards negative values of \(\sigma_z\otimes\sigma_z\).

With the model doing a good prediction of a few thousand random unitaries, we now take a look at real gates used in usual quantum computing.

\section{Results}
With the neural network trained with around 6000 points, we tried reproducing four one-qubit gates. We choose the known Hadamard, $X$, and $T$ gates (the latter is given by $Z^{1/4}$), as well as the operation where we keep the memory protected by a time interval $\tau$, which could be called the `Identity' operator. The steps are as follows: we get the predicted costate for the four gates from the neural network, then use \texttt{scipy.optimize.minimize} to get an ideal costate $\Lambda(0)$ for each gate. We then calculate the time evolution operator with Eq.~(\ref{unique}) and obtain the effective time-dependent Hamiltonian through
\begin{align}
    H\mleft(t\mright) = \mathcal{P}\left[ U(t) \Lambda\mleft(0\mright) U^\dagger\mleft(t\mright) \right].
\end{align}
Next, we use each resulting time-dependent Hamiltonian to solve the master equation, given by Eq.~(\ref{master}). For all the gates we choose $\left(\ket{0}+\ket{1}\right)/\sqrt{2}$ as the initial state. The reason is that a state of maximum superposition is the most affected by dephasing noise. So in the interaction picture, if the density operator keeps all of its coherence terms intact by the end of the evolution, this means the method of protection is working. 

Besides the evolution with the time-dependent Hamiltonian, we also solve the master equation with two other Hamiltonians to make comparisons. The first is with what we here call ``trivial Hamiltonian''. This Hamiltonian is defined as the constant Hamiltonian which would generate the target unitary should noise not be present. In practical terms, it is directly obtained through
\begin{align}
    H_{\text{triv}} = i \hbar \log{U\mleft(\tau\mright)}.
\end{align}
The second is the null Hamiltonian, that is, we solve the master equation with only noise acting on the system.

Since now we are analyzing the evolution of a density operator in the interaction picture, we quantify the fidelity, as a function of time, in terms of the initial density operator, that is,
\begin{align}
    F\mleft(t\mright) = \mathrm{Tr}\mleft[\rho\mleft(0\mright) \rho\mleft(t\mright)\mright].
\end{align}
This result follows immediately from the Uhlmann-Jozsa fidelity defined in Eq.~(\ref{UJfidelity}) when at least one of the density operators is pure, and in this case $\rho\mleft(0\mright)$ is always pure. Besides, since the analysis is made in the interaction picture, an ideally protected operation is identified as a density operator such that $\rho\mleft(0\mright) = \rho\mleft(\tau\mright)$, meaning $F\mleft(\tau\mright)=1$, while the noisy operation is identified with $\rho\mleft(\tau\mright)$ as a mixed state, meaning $F\mleft(\tau\mright) < 1$.
\begin{figure}
     \centering
     \includegraphics[width=0.482\textwidth]{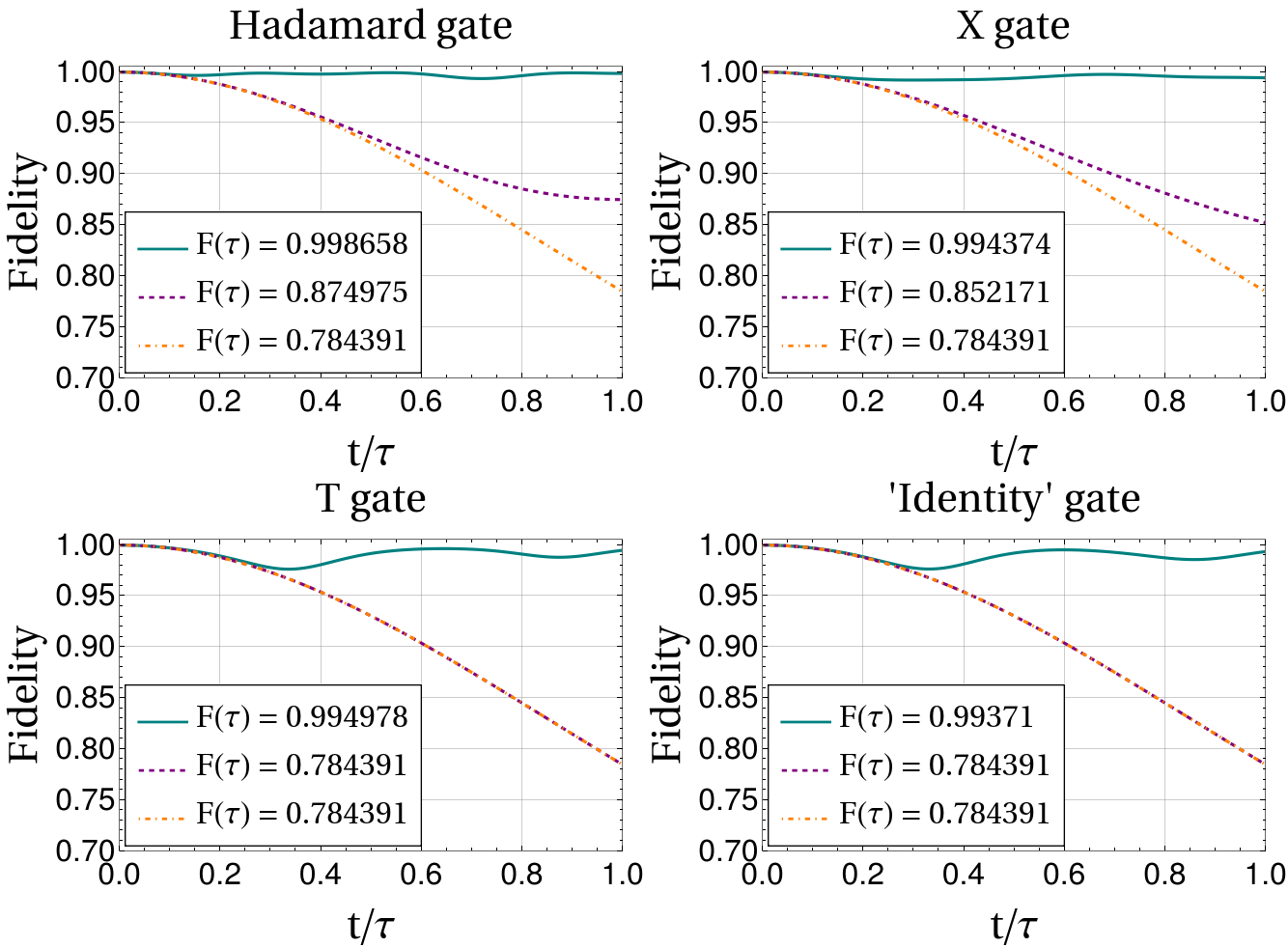}
        \caption{Fidelity as a function of time throughout the application of the quantum gates. The parameters chosen were $\eta=0.34$ and $\omega_c = \pi/(5\tau)$. The solid line is for the master equation solved with the optimal Hamiltonian, the dashed is for the trivial Hamiltonian and the dot-dashed is for the evolution due to noise exclusively. In the legends we show the fidelity at time $t=\tau$, that is, the gate fidelity.}
        \label{fidelities2}
\end{figure}

To attest the validity of our strategy, we suppose the dephasing environment with $\eta = 0.34$ and $\omega_c = \pi/\left(5 \tau\right)$. As was already mentioned, a longer correlation time, or equivalently, a lower $\omega_c$, places us in the CDD time regime, the ideal setting to test our optimal protective approach. The results for the fidelities of the four one-qubit gates are shown in Fig.~\ref{fidelities2}. Notice that the evolution due to noise exclusively, represented by the dot-dashed lines, is the same for all gates since the acting noise is identical in all cases.

As observed, the lowest gate fidelity obtained via optimal control was $0.99371$ for the Identity gate and the highest was $0.998658$, for the Hadamard gate. In comparison with the trivial Hamiltonian and the noise-free evolution, the fidelity by $t = \tau$ was bellow $0.8$ for the Identity and around $0.87$ for the Hadamard gate. This represents a significant improvement. For the $T$ gate, the evolution due only to noise coincides with the one using the trivial Hamiltonian. The reason is that since it is just a phase gate, the trivial Hamiltonian is a constant proportional to $\sigma_z \otimes I$, and this commutes with the noise operator, proportional to $\sigma_z \otimes \sigma_z$. Therefore, when treated in the perspective of the interaction picture, the two-time evolutions are identical. In other words, the noise is indifferent when applying only $\sigma_z \otimes I$ on the qubit. It is also worth mentioning that the trivial Hamiltonian for the Identity is just the null Hamiltonian since it corresponds to not applying any gates.

Our results indicate a significant improvement in gate fidelities using optimal control methods for continuous dynamical decoupling. This is in line with previous studies that have employed concatenated continuous driving to achieve robust dynamical decoupling, as demonstrated by Jianming et al.~\cite{Cai_2012}. Furthermore, our findings corroborate the efficacy of such techniques in practical scenarios, akin to the preservation of quantum spin coherence in biological environments as shown by Jianming et al.~\cite{PhysRevApplied.13.024021}.

\begin{figure}
        \includegraphics[width = 0.46\textwidth]{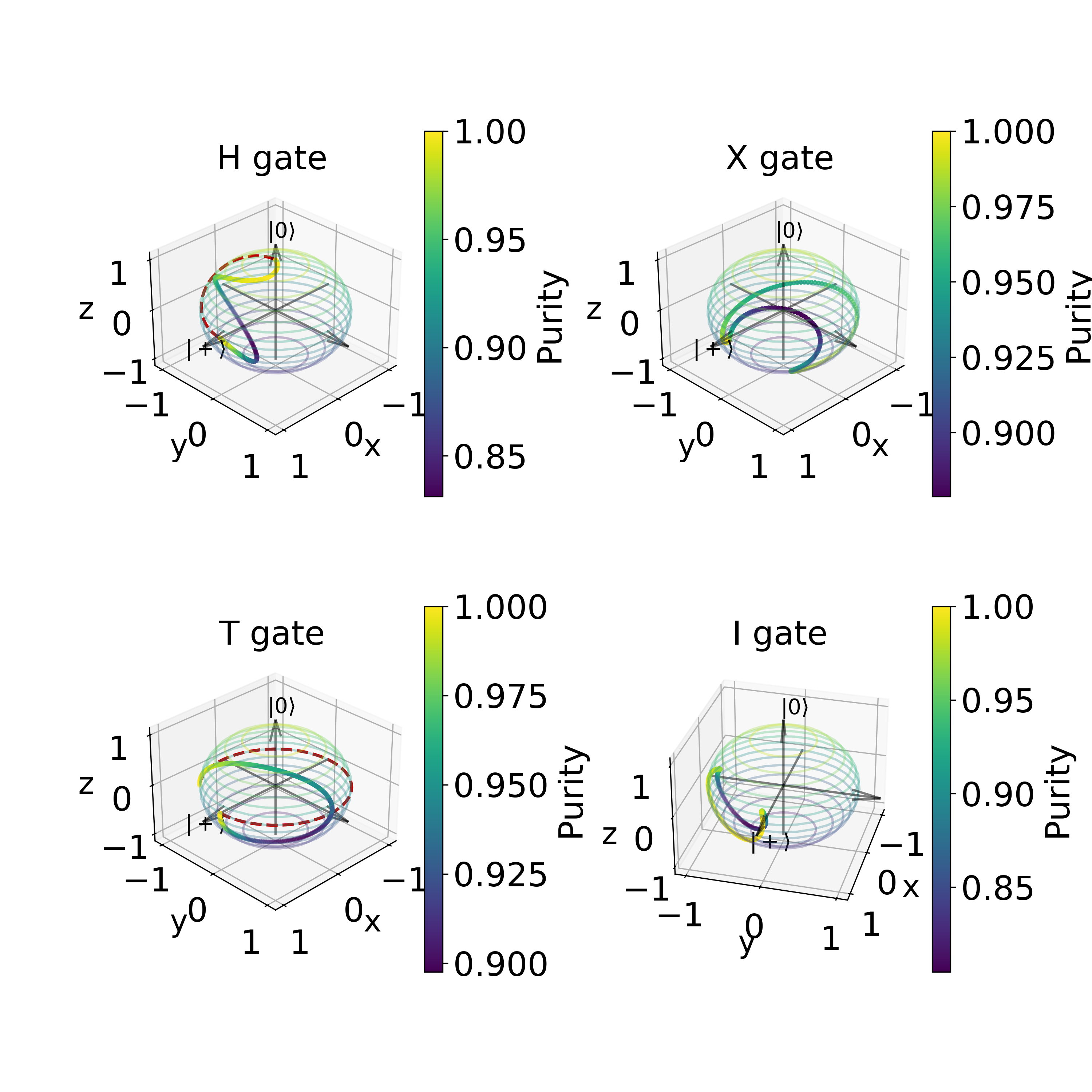}
        \caption{Evolution of the density matrix \(\rho(0)=\op{+}{+}\) with the application of four gates. Color represents purity. The evolution begins with a pure state, traverses inside the Bloch sphere, in which the purity is less than one, and emerges at its surface at \(t=\tau\), making sure that the gate mimics a unitary evolution. The dashed curves represent the ideal gate.}
        \label{fig:nicolas:rho-gates}
\end{figure}

We can check, for the gates present in Fig. \ref{fidelities2}, the evolution of the density matrix plotted on the Bloch sphere. As shown in Fig. \ref{fig:nicolas:rho-gates}, starting at a pure state, the qubit travels inside the Bloch sphere due to the noisy gates, and emerges at its surface when the computation is complete. Thus, it mimics a unitary transformation within a noise environment. We also compare it with the ideal evolution, plotted as a dashed line.

\begin{figure}
    \centering
    \includegraphics[width=0.482\textwidth]{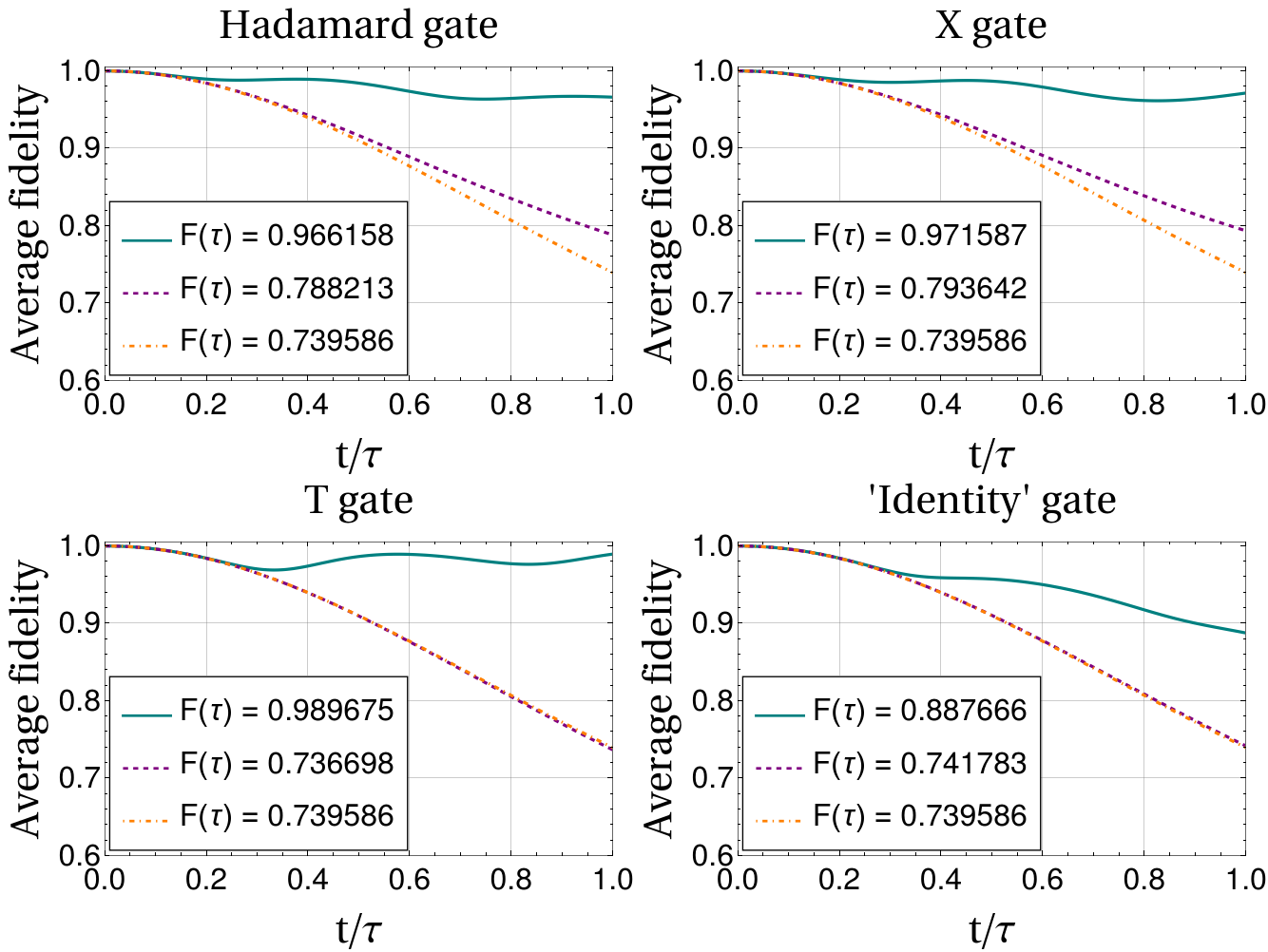}
        \caption{Average fidelities as a function of time the for gates, using the master equation with a Jaynes-Cummings interaction. The chosen parameters and the meaning of each line are the same as in Fig.~\ref{fidelities2}.}
        \label{avgfidelities}
\end{figure}

To check if the same control fields could be used for protection against amplitude damping noise, according to Eqs.~(\ref{fx})-(\ref{fz}), in the regime where $\omega_0 \approx \omega_c$, we also solved the master equation with a Jaynes-Cummings interaction, using the already known optimal control for dephasing with the axes permuted. The results are shown in Fig.~\ref{avgfidelities}. Differently from the graphs shown in Fig.~\ref{fidelities2}, we show the average fidelity, calculated as~\cite{bowdrey2002fidelity}
\begin{align}\label{avg_fidelity}
    \overline{F}(t) = \frac{1}{6} \sum_j \mathrm{Tr} \left[|j \rangle \langle j| \rho(t)\right],
\end{align}
where $|j\rangle$ runs over the six eigenstates of $\sigma_x$, $\sigma_y$ and $\sigma_z$. The reason is that, differently from dephasing noise, it affects all states instead of just the ones that start in a superposition state. As can be seen, the lowest gate fidelity obtained was $0.887666$ for the Identity gate and the highest was $0.989675$ for the $X$ gate. Although the gate fidelities are lower than the values shown in Fig.~\ref{fidelities2}, they still show a significant improvement over the unprotected case even though they were optimized primarily concerning dephasing noise. This result shows that it is possible to use the optimal fields for an interaction of the form $\sigma_z B(t)$ for other types of noise given that the control fields are appropriately rotated.

Now, it remains to verify how these obtained control fields compare, in energetic cost, with the fast-oscillating fields usually involved in the context of CDD, such as the methods described in~\cite{PhysRevA.75.022329, Napolitano2021protecting}.

According to Eq.~\ref{J}, we are calculating the energy cost for the implementation of a control Hamiltonian as
\begin{align}\label{energy_functional}
    \mathcal{E}(H, U) = \frac{1}{2} \int_0^\tau \mathrm{d}t\, \mathrm{Tr}\mleft\{ H(t) H(t) \mright\}.
\end{align}
So after obtaining the optimal control (OC) fields for the four chosen quantum gates, we calculated the required control fields for them in the context of general continuous dynamical decoupling (GCDD)~\cite{PhysRevA.75.022329, Napolitano2021protecting}, to compare the energetic cost between the two methods.

The motivation for this comparison is that, differently from the OC scheme, the Hamiltonian for the GCDD has the form $H(t) = H_\text{gate} + H_c(t)$, where the control $H_c(t)$ is a periodic operator whose period must be an integer multiple of the gate time and less than the noise correlation time, and $H_c(\tau) = I$. Therefore, since it is a method that involves rapidly oscillating fields, it is expected to be more costly, energetically speaking, than the optimal control scheme, and this is confirmed by the results shown in Table~\ref{tab}.

\begin{table}[h!]
    \centering
    \begin{tabular}{|r|c|c|c|c|}
         \hline
         & $H$ & $X$ & $T$ & $\mathbb{I}$ \\ \hline
         OC & \;16.8959\; & \;13.5178\; & \;13.6184\; & \;13.4660\; \\ \hline
         GCDD & \;368.102\; & \;396.018\; & \;384.992\; & \;394.784\; \\ \hline
    \end{tabular}
    \caption{Energetic cost for executing the four gates: Hadamard ($H$), $X$, $T,$ and the identity $\mathbb{I}$, using both methods: optimal control (OC) and general continuous dynamical decoupling (GCDD). All numeric values are in units of $\hbar/\tau$, meaning the actual energy is obtained by multiplying these values by $\hbar/\tau$.}
    \label{tab}
\end{table}

As can be seen, the energies differ in magnitude between $10$ and $100$ times for the chosen gates. This is numerical evidence that the optimal control for dynamical decoupling presented here is indeed more efficient than usual methods involving fast-oscillating fields. It is relevant to mention that such energetic costs for the CDD method correspond to the case of fast oscillating fields that generate fidelity values equivalent to the ones obtained with OC shown in Fig.~\ref{fidelities2}, that is, values between $0.99$ and $0.999$. It is possible, however, to obtain higher fidelities by increasing the oscillation period in CDD and thus, the energy cost. Therefore, this comparison does not lead to the conclusion that OC is strictly superior to CDD in every scenario, but only when considering that the obtained values of gate fidelity between $0.99$ and $0.999$ are high enough.

\begin{figure}
     \centering
     \includegraphics[width=0.482\textwidth]{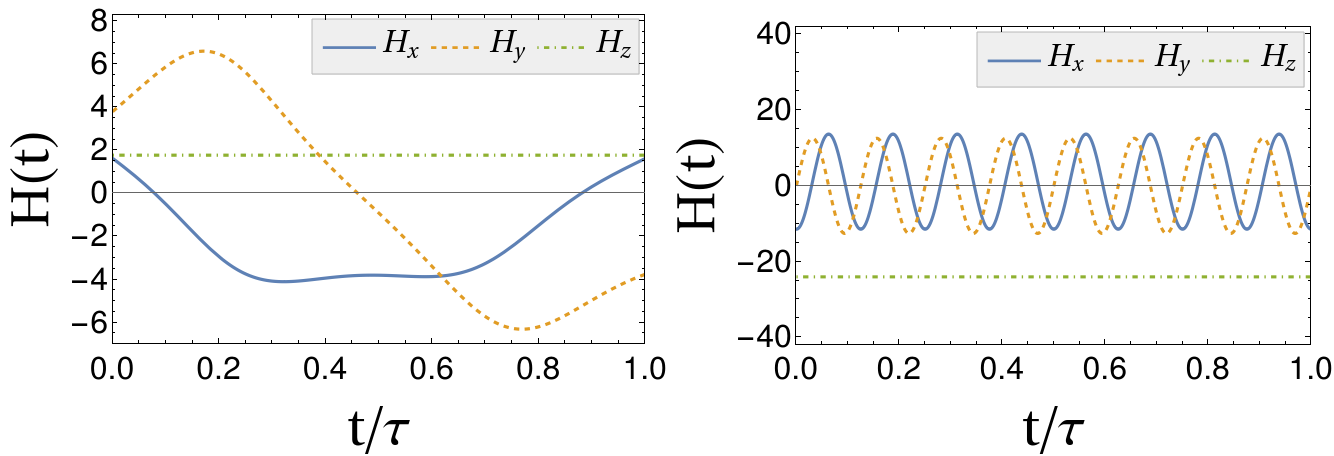}
        \caption{Components $\sigma_x$, $\sigma_y$ and $\sigma_z$ of the control Hamiltonian for both, optimal control (left) and general continuous dynamical decoupling (right), for the Hadamard gate. The energy values in the y-axis are given in units of $\hbar/\tau$, meaning the actual energy is obtained by multiplying these numerical results by $\hbar/\tau$.}
        \label{components_H}
\end{figure}

\begin{figure}
    \centering
    \includegraphics[width=0.47\textwidth]{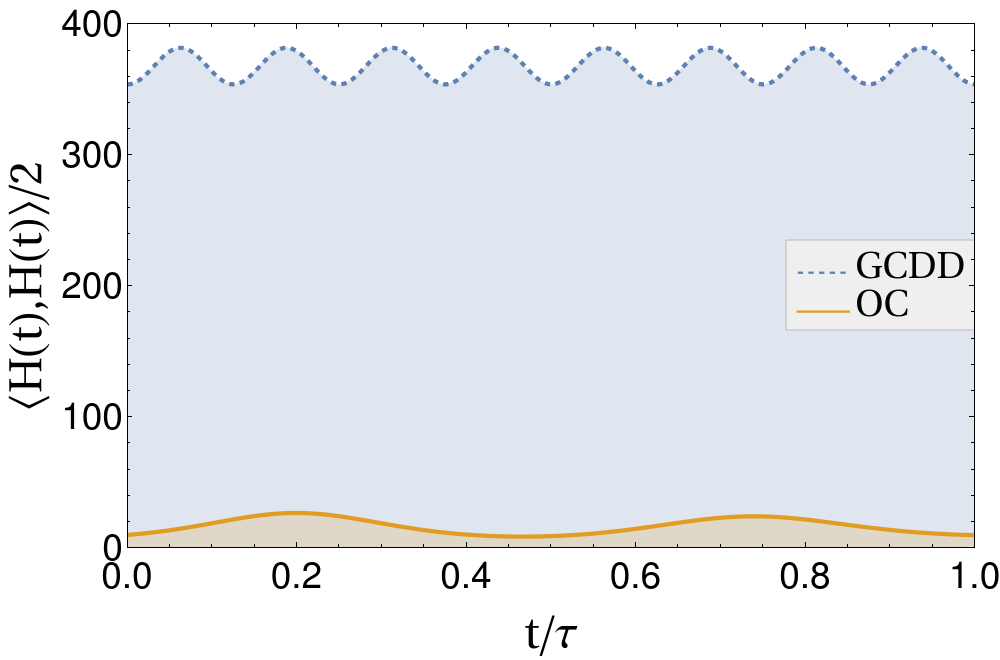}
    \caption{Time-dependent energy function $\mathrm{Tr}\mleft\{ H(t) H(t) \mright\}/2$ for both, optimal control and general continuous dynamical decoupling. The total energy cost in units of $\hbar/\tau$, according to Eq.~\ref{energy_functional}, is given by the areas under the curves.}
    \label{comparison}
\end{figure}

To compare the typical time profile dependence of the Hamiltonians we show, in Fig.~\ref{components_H}, the three components of the control for the Hadamard gate in both methods: OC and GCDD. We also show the comparison in energy cost between the two methods in Fig~\ref{comparison}, by plotting the energy function $\mathrm{Tr}\mleft\{ H(t) H(t) \mright\}/2 = \left[H_x(t)^2 + H_y(t)^2 + H_z(t)^2\right]/2$. Besides confirming an improvement concerning energy consumption, this figure, along with Fig.~\ref{components_H}, shows that the derivatives of the control fields of OC are considerably smaller than the ones in GCDD. This suggests a more efficient regime for field variations, that is, less imprecision when turning the field on/off, as well as when changing the field to the configuration of another quantum gate when applying several in sequence.

%We then verified how the fidelities would change if we set the parameters such that we increase the bath correlation time as well as the amplitude $\eta$. 

%The chosen numerical values for this case were $\eta = 0.34$ and $\omega_c = \pi/\left(5 \tau\right)$ that is, the cutoff frequency is $10$ times lower and a noise amplitude $34$ times higher. The reason we chose the factor of $34$ for the noise amplitude is that, with these parameters, the fidelity in the unprotected case is around $0.784$, which is about the same as the previous case. So in the time scale of $\tau$, both noises are equally destructive in the absence of control. These results are shown in Fig.~\ref{fidelities2}.

% Compared with the previous case, the optimal Hamiltonian estimated by the neural network is considerably more efficient in this regime. This is a numerical confirmation of the fact that this protocol of optimal control is better suited for noise with a long correlation time compared to the gate time.

\color{black}
\section{Conclusion}

We present an optimal control strategy designed for the CDD regime where the noise correlation time is long compared to the time scale in which the control fields can vary~\cite{9835639,PhysRevA.91.032325,Dong2016,Li2022pulselevelnoisy,Silverio2022pulseropensource,9951219}. In such situation, we use the time-dependent purification scheme we envisaged to emulate the noise through a unitary evolution, by introducing an auxiliary qubit to be traced over after the optimization is completed. We use a sub-Riemannian geometry framework for the two-qubit unitary group, to show that we can, in principle, efficiently implement any single-qubit gates despite the presence of dephasing noise.

Recently, Koch and colleagues have addressed the purification and reset problem for a qubit using optimal control techniques, providing highly relevant insights into the problem of optimized control fields for continuous dynamical decoupling. Their methodology demonstrates significant advancements in qubit reset protocols, which aligns closely with our approach to employing auxiliary qubits for implementing purification schemes~\cite{Koch2021}.

Our control fields, represented by continuous functions of time, can be used at the pulse level in the coding structure of current available quantum computers. Naturally, as exposed above, our optimal control requires previous spectroscopy of the environmental noise, which is obtainable using different techniques, including Ramsey spectroscopy and pulsed dynamical decoupling~\cite{Szankowski2017, PhysRevLett.107.230501, PhysRevApplied.15.014033, PhysRevLett.107.170504, Almog_2011, Burgardt_2023, PhysRevLett.126.250507, PhysRevLett.111.093604}.

%Using our strategy, we derive the optimized time-dependent control Hamiltonian that produces gates with notably high fidelities. 
%We investigated how well we can translate the optimized control Hamiltonian in the unitary dynamics to the original master equation. Our results show that it is a promising technique, even if as a preliminary step to be refined afterward. 

Also, we train a neural network to efficiently estimate control Hamiltonians for any on-demand target unitaries covering the $\mathrm{SU}\mleft(2\mright)$ unitary group. We show that, after having trained the neural network to guess the optimal control for a set of parameters, it can be used to guess the results for a new set of parameters, albeit sufficiently close to the previous ones, so that ulterior refinement produces corresponding high-fidelity optimal gates for the new parameters. Previous works using machine learning to enhance the efficiency of the production of on-demand quantum circuits~\cite{Perrier_2020, SWADDLE20173391, Swaddle-dissertation, Youssry2020} were mainly concerned with the whole quantum circuit, considering each quantum gate as an instantaneous and ideal operation on the qubits on which they acted. Thus those studies were not focusing on the pulse level for achieving optimal CDD.

Our method of using quantum optimal control can be seen as a proof of principle for the efficient design of high-fidelity quantum circuits. As mentioned in the introduction, a single entangling two-qubit gate combined with on-demand single-qubit gates is all one needs to form the basic modules of a complex quantum circuit. 
Our present approach is to be considered as another tool in the whole toolbox to address an improved treatment of quantum algorithms over the current dominant NISQ framework. 

In our ongoing investigations into this topic, we intend to devise an analogous method to obtain optimized control Hamiltonians rendering on-demand two-qubit gate operations. This problem involves a substantial increase in the number of dimensions to be considered. Our difficulty will be in generating a sufficient number of training solutions to obtain a neural network capable of augmenting the training data as we have done here in the single-qubit case.

\begin{acknowledgments}
R.d.J.N. and F.F.F acknowledges support from Funda\c{c}\~ao de Amparo \`a Pesquisa do Estado de S\~ao Paulo (FAPESP), project number 2018/00796-3 and 2023/04987-6, and also
from the National Institute of Science and Technology for Quantum
Information (CNPq INCT-IQ 465469/2014-0) and the National Council
for Scientific and Technological Development (CNPq). N. A. da C. M.
acknowledges financial support from Coordena\c{c}\~ao de Aperfei\c{c}oamento
de Pessoal de N\'ivel Superior (CAPES), project number 88887.339588/2019-00. A. H. da S. acknowledges financial support from Conselho Nacional de Desenvolvimento Científico e Tecnológico (CNPq), project number 160849/2021-7. We also thank Prof. Leonardo Kleber Castelano for a careful reading of the manuscript and for pointing out some typographical errors. F.F.F. acknowledges support from ONR, Project No. N62909-24-1-2012.
\end{acknowledgments}

\section{Appendix}
\subsection{Equivalence between Eqs.~(\ref{master}) and (\ref{effective master}) when $S\mleft(t^{\prime}\mright)  \approx  S\mleft(t\mright)$}\label{appA}
It follows from Secs.~\ref{model}, \ref{qoc}, and Eqs.~(\ref{master}) that when $S\mleft(t^{\prime}\mright)  \approx  S\mleft(t\mright)$ the master equation yields
\begin{align}
    \frac{\mathrm{d}\rho_{IS}\mleft(t\mright)}{\mathrm{d}t} = \frac{\dot{\mu}\mleft(t\mright)}{ 2 \mu\mleft(t\mright)} \left[  \rho_{IS}\mleft(t\mright) - S\mleft(t\mright) \rho_{IS}\mleft(t\mright)  S\mleft(t\mright) \right].
    \label{mastermucontrolintpic}
\end{align}
Taking Eqs.~(\ref{h(t)})-(\ref{drift_hamiltonian}) into account when manipulating Eq.~(\ref{effective master}) in this regime and partially tracing over the auxiliary qubit, we obtain
\begin{align}
    \dfrac{\mathrm{d} \rho_{IS}^{(e)}\mleft(t\mright)}{\mathrm{d}t} = \frac{-2h\mleft(t\mright)}{ \hbar^{2}}  \int_{0}^{t} \mathrm{d}t^{\prime} \, h\mleft(t^{\prime}\mright) \mathscr{D}_{S\mleft(t\mright)}\big(  \rho_{IS}^{(e)}\mleft(t^{\prime}\mright)\big),  \label{rho pure effective master D}
\end{align}
where we define
\begin{align}
    \mathscr{D}_{S\mleft(t\mright)}(A) = A -  S\mleft(t\mright) A S\mleft(t\mright).  \label{D calligraphic}
\end{align} 
By iteration to all orders, Eq.~(\ref{rho pure effective master D}) can be summed up and we obtain
\begin{align}
  \dfrac{\mathrm{d} \rho_{IS}^{(e)}\mleft(t\mright)}{\mathrm{d}t} = \frac{\dot{\mu}\mleft(t\mright)}{2\mu\mleft(t\mright)}  \left[ \rho_{IS}^{(e)}\mleft(t\mright) -  S\mleft(t\mright) \rho_{IS}^{(e)}\mleft(t\mright)S\mleft(t\mright) \right] ,\label{drho/dt effective intpic proof}
\end{align} which is equivalent to Eq.~(\ref{mastermucontrolintpic}).

\subsection{Equivalence between Eqs.~(\ref{master}) and (\ref{effective master}) when $\omega_c t \ll 1$ and \ref{approx B} is valid}\label{appB}

Here we show that the condition $\omega_c t \ll 1$ leads to the master equation evolution, Eq.~(\ref{master}), to coincide approximately with the effective Hamiltonian evolution, Eq.~(\ref{effective master}). For long correlation times, which implies small values of $\omega_c$, we see from Eq.~(\ref{decaying coherence}) that $\mu\mleft(t^{\prime}\mright)\approx\mu\mleft(t\mright)$ so that $h\mleft(t^{\prime}\mright)\approx h\mleft(t\mright)$ within the integral of Eq.~(\ref{effective master}). Because a long correlation time of the boson bath means that if $t\ll t_{c},$ then \ref{approx B} is valid within the integral of Eq.~(\ref{master}), and it follows that
\begin{align}
[h\mleft(t\mright)]^{2} &\approx \mathrm{Tr}_{B}\mleft[B\mleft(t\mright)B\mleft(t\mright)\rho _{B}\mleft(0\mright)\mright] \nonumber \\
&\approx \hbar^{2}\eta \omega_{c}^{2}+2\hbar^{2}\eta \omega_{T}^{2}\psi^{(1))}\mleft(1+\frac{\omega_{T}}{\omega_{c}}\mright) \nonumber \\
&\equiv \hbar^{2}\lambda_{0}.
\end{align}
Therefore, both Eqs.~(\ref{master}) and (\ref{effective master}) now write approximately the same, namely,
\begin{align*}
\frac{\mathrm{d}\rho_{IS}\mleft(t\mright)}{\mathrm{d}t} &\approx -\lambda_{0}\int_{0}^{t}\mathrm{d}t^{\prime}\,\left[S\mleft(t\mright),\left[S\mleft(t^{\prime}\mright),\rho_{IS}\mleft(t\mright)\right]\right],
\end{align*}
up to the second order in the noise strength $\eta$.

\bibliography{references.bib}

\end{document}